\documentclass{article}

\usepackage{arxiv}

\usepackage[utf8]{inputenc}
\usepackage[T1]{fontenc}
\usepackage{url}
\usepackage{booktabs}
\usepackage{amsfonts}
\usepackage{amssymb}
\usepackage{nicefrac}
\usepackage{microtype}
\usepackage{graphicx}
\usepackage[round]{natbib}
\usepackage{amsmath}
\usepackage{bbm}
\usepackage{algorithm}
\usepackage{amsthm}
\usepackage[hidelinks]{hyperref}

\newtheorem{theorem}{Theorem}

\newtheorem{lemma}{Lemma}

\title{Sequential Probability Ratio Test using Z-Statistics (SPRT-z): A Practical Approach for Online Experimentation}

\author{
    Derek L. Ho \\
    Principal Data Scientist \\
    Atlassian \\
    Sydney, Australia \\
    \texttt{dho@atlassian.com}
    \and
    Emma G. Thomas \\
    Senior Data Scientist \\
    Atlassian \\
    Los Angeles, United States \\
    \texttt{ethomas2@atlassian.com}
}

\renewcommand{\shorttitle}{SPRT-z: Practical Online Experimentation}



\begin{document}
\maketitle

\begin{abstract}
Modern online experimentation platforms produce data continuously, yet practitioners often apply Fixed Horizon Testing (FHT) while repeatedly peeking at results, inflating Type I error. Popular always-valid methods permit early efficacy stopping but do not natively combine futility stopping, business-relevant launch thresholds, and Type II error control. We adapt Wald's Sequential Probability Ratio Test (SPRT) to meet these needs. SPRT-z uses a large-sample normal approximation to Hajnal's sequential $t$-test, reducing computation and enabling two Brownian methods. Scale-Free Horizon Calibration (SFHC) sets a maximum sample size to approximately recover fixed-horizon power under discrete monitoring and futility stopping. A Brownian median unbiased estimator (MUE) and confidence intervals are defined for every stopping outcome. A validity analysis provides exact Gaussian-oracle accounting for the deployed stopping rule and a qualified fixed-look approximation for stationary finite-variance outcomes.

In simulations with identically distributed arms, SPRT-z kept Type I error at or below nominal, tracked fixed-horizon power within a few percentage points except under the coarsest monitoring, and reduced expected sample size in all but one underpowered setting. The MUE reduced early-stopping bias, with close-to-nominal interval coverage in most settings; we report exceptions and operating rules. Stress tests identify outcome shapes and monitoring schedules that require separate error and power calibration. Across production experiments at Atlassian, median sample saving relative to fixed-horizon plans was 45.5\%. More than 75\% of decisions came from the futility boundary, supporting its practical importance.
\end{abstract}


\section{Introduction}

Modern experimentation platforms expose results continuously, but experimenters often apply Fixed Horizon Testing (FHT) as if they had observed the data only once. Acting on these interim results without adjustment (``peeking'') inflates Type I error. FHT also requires experiments to run to a predetermined sample size even when the treatment effect is clearly flat.

Sequential tests address these problems by allowing early decisions while accounting for interim monitoring. Group Sequential Testing (GST) controls error over a finite schedule of analyses \citep{jennison1999group}. Always Valid Inference (AVI) permits unlimited peeking, but has less power than GST at a fixed sample size \citep{johari2022}. Online A/B tests motivate frequent monitoring, which makes the distinction between an exact always valid reference and an approximate deployed statistic consequential.

Wald's Sequential Probability Ratio Test (SPRT) is an early always-valid test \citep{wald1947}. It compares simple hypotheses, $H_0: \delta = \delta_0$ and $H_1: \delta = \delta_1$, and can stop early for efficacy or futility. Its simple hypotheses are often considered restrictive: when the true effect lies between $\delta_0$ and $\delta_1$, the statistic can remain between the boundaries longer than an FHT would run.

Modern alternatives include mixture SPRT (mSPRT) \citep{johari2022}, the Generalization of Always Valid Inference (GAVI) \citep{howard2021}, and E-value methods \citep[e.g.,][]{ramdas2023}. \cite{koning2026} constructs tests that allow early stopping without sacrificing fixed-horizon power by randomizing interim decisions. The Bayesian method of \cite{dengLuChen2016} uses posterior odds under a calibrated prior instead of frequentist alpha spending. Like the other AVI methods above, it does not natively control Type II error or support futility stopping.

We revisit Wald's SPRT because large-scale experimentation benefits from early efficacy and futility stopping, business-relevant launch thresholds, and explicit targets for Type I and Type II error.\footnote{The method is deployed at Atlassian; Section~\ref{sec:production} reports its measured performance.}

A two-boundary GST can provide all three, but only with a prespecified maximum information and interim schedule independent of the observed effect. Extending the test or timing analyses around a promising trend invalidates its Type I guarantee without correction. By contrast, an ideal SPRT likelihood-ratio martingale retains its boundary validity for any realized sequence of looks; overruns, truncation, and unplanned peeking affect power but not Type I error. The prequential statistic deployed here is not an exact martingale, so we instead treat it as a finite-horizon, cadence-qualified procedure. Section~\ref{sec:validity} compares it with a Gaussian oracle and states the conditions under which that comparison informs Type I error. Its boundaries enable efficacy and futility stopping against business-relevant simple hypotheses, while SFHC approximately calibrates Type II error at the design MDE (Section~\ref{sec:horizon}).

Popular AVI methods bound Type I error over an infinite horizon and stop early for efficacy, but do not natively combine futility stopping, an MDE-based launch criterion, and Type II error control. A confidence sequence can support futility stopping by excluding a threshold, but does not also guarantee power at a specified non-null effect. The method of \cite{koning2026} can provide all three properties, but requires randomized interim decisions that may be unacceptable to decision makers.

Futility stopping preserves traffic and accelerates iteration. Because 70--90\% of shipped features on major platforms fail to produce a significant positive effect \citep{kohavi2020trustworthy}, stopping futile tests may save more capacity than efficacy stopping. Futility accounted for most decisions in our deployment (Section~\ref{sec:production}).

Explicit Type II error control lets teams choose the probability of detecting a business-relevant effect and trade that power against run time.

For this use case, SPRT's simple alternative is a practical threshold: \textit{What minimum uplift, or maximum acceptable regression, justifies launch?} Under continuous, untruncated monitoring, Wald's SPRT minimizes expected sample size at the two design hypotheses among tests with the same error rates \citep{wald1947}. Batch monitoring and truncation reduce this advantage; Section~\ref{csd:simresult} measures the remaining savings.

Our implementation adapts Hajnal's sequential $t$-test \citep{hajnal1961} by replacing its costly exact $t$ calculation with a large-sample normal approximation, which we call SPRT-z.

Scale-Free Horizon Calibration (SFHC) caps the long run times that occur between $H_0$ and $H_1$ while approximately recovering the desired power at the business threshold. A related simulation procedure estimates a Median Unbiased Estimator (MUE) and confidence interval (CI) to correct early-stopping bias.

The four contributions address computation, planning, estimation, and validity:
\begin{enumerate}
    \item \textbf{SPRT-z}: a normal approximation to Hajnal's SPRT-t for large online experiments.
    \item \textbf{SFHC}: a variance-independent Monte Carlo bisection on the standardised $Z$-scale that calibrates a maximum sample size under discrete monitoring and futility stopping.
    \item \textbf{Brownian MUE and confidence intervals}: a six-region stagewise ordering for efficacy, futility, and horizon-truncated outcomes, with empirical assessment of bias and coverage.
    \item \textbf{Validity analysis}: exact Gaussian-oracle accounting for the deployed stopping rule, a fixed-look local limit under stationary finite-variance sampling, and simulation audits. Seventeen stress scenarios compare the deployed form with fixed-horizon, group sequential, and mSPRT tests on identical streams (Appendix~\ref{app:comp_attack}).
\end{enumerate}

Sections~\ref{sec:core_mechanic} and~\ref{sec:prequential} develop the statistic and relative-MDE updates. Sections~\ref{sec:horizon} and~\ref{sec:validity} address planning and validity; Sections~\ref{sec:empirical} and~\ref{sec:bias} evaluate testing and estimation; Section~\ref{sec:production} reports deployment; and Section~\ref{sec:conclusion} concludes. Appendices provide validity proofs (Appendix~\ref{app:compensator}), audits and stress tests (Appendix~\ref{app:validity_audits}), efficiency comparisons (Appendix~\ref{app:comparisons}), SFHC and MUE mechanics (Appendices~\ref{app:sfhc} and~\ref{app:mue}), and a Bernoulli replication (Appendix~\ref{app:simulation_details}).

\section{The Core Mechanic: SPRT-z on Matured Cross-Sectional Data}
\label{sec:core_mechanic}

We formulate SPRT-z for fixed-window cross-sectional metrics, where each step adds a disjoint cohort of matured units (Section~\ref{subsec:fixedobswindow}). We then explain the computational motivation (Section~\ref{subsec:computational}) and derive the statistic and boundaries (Sections~\ref{subsec:formulation}--\ref{subsec:boundaries}).

\subsection{Enforcing Independent Increments via Fixed Observation Windows}
\label{subsec:fixedobswindow}

Sequential theory requires each monitoring step to add data independent of earlier steps. Continuously observed users instead generate autocorrelated longitudinal panels. We therefore assign each unit a \textit{fixed observation window} and admit its metric only after that window closes.

For a unit $i$ enrolled at time $E_i$, we compute its final metric $X_i$ over $[E_i,E_i+W]$ (e.g., seven days) and discard later telemetry.

The test admits unit $i$ only when $t \ge E_i+W$. Each step thus adds a disjoint cohort of matured users. Independence of its mean-difference increments additionally requires independent outcomes across units and cohorts; disjoint windows alone do not exclude common shocks or interference. Hereafter, \emph{cohort} means the treatment and control units whose windows close at the same step; treatment and control populations are called groups or arms.

\subsection{Computational Limitations of SPRT-t}
\label{subsec:computational}

Hajnal's exact SPRT-t evaluates a non-central $t$ likelihood ratio, or a non-central $F$ ratio for two-sided tests \citep{hajnal1961}. We benchmarked 10,000 concurrent metric updates over 1,000 repetitions in R, comparing \texttt{dnorm} with \texttt{dt} and \texttt{df} using noncentrality parameters and the \texttt{microbenchmark} package. The non-central $t$ and $F$ densities were approximately 38 and 15 times slower than the standard normal density; absolute timings depend on platform and interpreter. These distributions also require numerical approximation, whereas the normal density has a simple closed form. The normal approximation enables the Brownian calibration and bias correction in Sections~\ref{sec:horizon} and~\ref{sec:bias}.

\subsection{The SPRT-z Formulation and Burn-in Period}
\label{subsec:formulation}

At the large samples typical of online experiments, the $t$ distribution approaches the standard normal. SPRT-z therefore operates on the cumulative $z$-score:

\begin{equation}
    \label{eq:zscore}
    z_{m} = \frac{\bar{x}_{t,m} - \bar{x}_{c,m}}{\sqrt{\frac{s_{t,m}^2}{n_{t,m}} + \frac{s_{c,m}^2}{n_{c,m}}}}
\end{equation}
where $\bar{x}_{g,m}$ and $s^2_{g,m}$ are the sample mean and variance for group $g\in\{t,c\}$, and $n_{g,m}$ is its cumulative sample size at step $m$.

Replacing the unknown population variances $\sigma_t^2$ and $\sigma_c^2$ with unpooled sample variances $s_{t,m}^2$ and $s_{c,m}^2$ makes the normal likelihood approximation rather than exact. Section~\ref{sec:validity} states the additional conditions needed for an asymptotic error result.

Implementations should therefore require a burn-in period before monitoring or making decisions, with a minimum sample in both groups (e.g., $n_{min} \ge 100$ per group) and sufficient degrees of freedom for the $t$ and $Z$ distributions to be close ($\nu \ge 200$). High-traffic experiments typically accumulate these 200 units within minutes or hours. These thresholds are operational checks, not a finite-sample error guarantee for skewed outcomes or unstable variances.

After burn-in, we model the cumulative score as
\begin{equation}
    z_{m} \sim N(\psi_{m}, 1)
\end{equation}
The likelihood ratio compares the alternative and null densities, using
\begin{equation}
    f(z_m; \psi_m) = \frac{1}{\sqrt{2\pi}} e^{-\frac{1}{2}(z_m - \psi_m)^2}
\end{equation}
The non-centrality parameter $\psi_m$ represents accumulated signal at the absolute Minimum Detectable Effect $\delta_{MDE}$:
\begin{equation}
    \psi_m = \frac{\delta_{MDE}}{\sqrt{\frac{\sigma_t^2}{n_{t,m}} + \frac{\sigma_c^2}{n_{c,m}}}}
\end{equation}

\subsection{Log-Likelihood Ratio}
\label{subsec:llr}
At each step $m$, SPRT-z uses a cumulative log-likelihood-ratio score $C_m$ to decide whether to stop or continue. We write $C_m^+$ and $C_m^-$ for the two directional scores and $C_m^{(2)}$ for the two-sided score.

\textbf{For a One-Sided Test:}
The directional scores compare the alternative mean $+\psi_m$ or $-\psi_m$ with the zero-effect null:
\begin{equation}
    C_m^\pm = \ln \frac{f(z_m; \pm\psi_m)}{f(z_m; 0)} \label{eq:llr_one_sided}
\end{equation}
Substituting the normal density gives
\begin{equation}
    C_m^\pm = \pm z_m\psi_m - \frac{1}{2}\psi_m^2
\end{equation}

\textbf{For a Two-Sided Test:}
The alternative mixes effects $+\psi_m$ and $-\psi_m$ with equal weights. Wald's weight-function method \citep{wald1947} gives the standard symmetric two-point mixture \citep{robbins1970, lai1976, siegmund1985sequential}:
\begin{equation}
    C_m^{(2)} = \ln \left( \frac{0.5 \cdot f(z_m; +\psi_m) + 0.5 \cdot f(z_m; -\psi_m)}{f(z_m; 0)} \right)
\end{equation}

Substituting the normal density and $\cosh(x)=(e^x+e^{-x})/2$ gives
\begin{equation}
    \label{eq:llr_two}
    C_m^{(2)} = \ln(\cosh(z_m\psi_m)) - \frac{1}{2}\psi_m^2
\end{equation}

With independent Gaussian observations, known arm variances, a fixed target, and exactly proportional arm counts at every look, the scores evaluated using true variances are Gaussian log-likelihood ratios \citep{wald1947}. The exponentiated directional scores and their fixed equal-weight mixture are then valid test martingales. SPRT-z instead updates its target and estimates unpooled variances. Section~\ref{sec:validity} gives an exact decomposition of the resulting score gap and a local fixed-look approximation under separate stationary assumptions.

\subsection{Decision Boundaries and Termination}
\label{subsec:boundaries}

Wald's upper and lower boundaries for Type I and Type II inputs $\alpha$ and $\beta$ are \citep{wald1947}
\begin{equation}
    A = \ln\left(\frac{1 - \beta}{\alpha}\right) \quad \text{and} \quad B = \ln\left(\frac{\beta}{1 - \alpha}\right)
\end{equation}
The 50/50 mixture already accounts for the two-sided alternative, so $\alpha$ is not divided by two.

For a valid likelihood-ratio reference, the efficacy boundary $A$ gives a time-uniform Ville bound of $e^{-A}=\alpha/(1-\beta)$ (0.0625 when $\alpha=0.05$ and $\beta=0.20$). Control at $\alpha$ with the Wald boundary also depends on the two-boundary stopping rule and overshoot. Setting $A=\ln(1/\alpha)$ gives a 5\% Ville bound for the valid reference, but the deployed score retains an oracle gap at either boundary. Section~\ref{sec:validity} and Appendix~\ref{app:compensator} analyse this distinction; all reported rates use the shipped, truncated two-boundary design.

These boundary inputs do not specify realised error rates for the deployed rule. We measure those rates under the tested outcomes and schedules. At the shipped inputs $(0.05,0.20)$, Type I error was below 0.05 in the identically distributed-arm main study (Table~\ref{tab:validation}); Appendix~\ref{app:comp_matched} finds that $a=0.077$ realises 5\% on the stationary generator. The broader stress battery and its recalibration are reported in Appendix~\ref{app:comp_attack}. Deployments should calibrate inputs for the metric family, cadence, and horizon, and publish the measured rates.

Let $C_m$ denote the score for the chosen directional or two-sided test. The decision rule at step $m$ is
\begin{equation}
    \text{Decision} = 
    \begin{cases} 
    \text{Accept } H_1, & \text{if } C_m \ge A \\
    \text{Accept } H_0, & \text{if } C_m \le B \\
    \text{Continue}, & \text{if } B < C_m < A
    \end{cases}
\end{equation}

\section{Handling Relative MDEs via Prequential Updates}
\label{sec:prequential}

Online A/B tests often target a relative rather than absolute change. Given control mean $\mu_c$, the relative uplift $r_{MDE}$ corresponds to $\delta_{MDE}=\mu_c r_{MDE}$.

Historical estimates of $\mu_c$ may be unavailable, stale, seasonal, or incorrectly sourced. We instead update it from the experiment.

A \textit{prequential} update uses the control mean $\bar{x}_{c,m-1}$ and unpooled variances $s_{t,m-1}^2,s_{c,m-1}^2$ from units matured before step $m$:
\begin{equation}
    \label{eq:prequential_psi}
    \psi_m = \frac{\bar{x}_{c,m-1} \cdot r_{MDE}}{\sqrt{\frac{s_{t,m-1}^2}{n_{t,m}} + \frac{s_{c,m-1}^2}{n_{c,m}}}}
\end{equation}

The lag makes $\psi_m$ measurable with respect to the historical filtration $\mathcal{F}_{m-1}$, but predictability of the target does not make the deployed cumulative score a martingale. The score re-evaluates all matured observations at the current target and estimated scales. Section~\ref{sec:validity} compares it with an oracle Gaussian reference using a fixed target and true variances; Appendix~\ref{app:anatomy_proof} separates their gap into target and scale components.

\section{Sample Size Planning and Maximum Horizons}
\label{sec:horizon}

An exact test martingale remains valid at any stopping time and therefore needs no fixed sample size \citep{johari2022}. For deployed SPRT-z, Section~\ref{sec:validity} analyses this property rather than assuming it. Traffic planning and rollout schedules may nevertheless require a maximum duration.

Wald's identity gives the Expected Sample Size, or Average Sample Number (ASN), under $H_0$ or $H_1$ \citep{wald1947}; it does not specify when to truncate an inconclusive test.

Moreover, Wald's derivation assumes an update after every observation, whereas platforms commonly monitor in daily batches.

We recommend one of two maximum-horizon rules, depending on telemetry quality.

\subsubsection*{Paradigm 1: Business-Driven Maximum Horizons}

With volatile or unreliable telemetry, set the horizon from business cycles and seasonality, such as 28 days for four weekly cycles. Truncate inconclusive tests at that horizon and default to $H_0$.

\subsubsection*{Paradigm 2: Scale-Free Horizon Calibration (SFHC)}
With stable baselines and predictable daily traffic $n_{daily}$, \textbf{Scale-Free Horizon Calibration (SFHC)} connects fixed-horizon planning to sequential execution. Appendix~\ref{app:sfhc} gives the algorithm and pseudocode (Algorithm~\ref{alg:sfhc}).

SFHC approximates cumulative evidence under $H_1$ by Brownian motion with drift. A Monte Carlo search inflates the fixed-horizon sample size to recover power lost to discrete monitoring and futility stopping:

\begin{enumerate}
    \item \textbf{FHT anchor:} Compute $N_{FHT}$ from the baseline variance, target MDE $\delta$, and error inputs $(\alpha,\beta)$.
    \item \textbf{Brownian simulation:} Use $N_{FHT}$ and $n_{daily}$ to define a monitoring grid, then simulate thousands of standardised Brownian paths drifting toward the FHT endpoint ($Z_{FHT}$), against the SPRT-z efficacy and futility boundaries.
    \item \textbf{Bisection:} Increase $N_{max}$ until the truncated sequential design approximately recovers the anchor's power.
\end{enumerate}

SPRT-z is monitored until $N_{max}$; a test still in the continuation region then stops and accepts $H_0$. Section~\ref{csd:simresult} measures the realised error rates, while Section~\ref{sec:validity} analyses the rule on a specified finite monitoring grid. The horizon also gives stakeholders a maximum completion date.

\section{Type I Error under Prequential Updating and Batch Monitoring}
\label{sec:validity}

We analyse the deployed Type I error by comparing the SPRT-z score $C_m$ with its oracle Gaussian reference $C_m^*$. Section~\ref{subsec:compensator_main} gives exact rejection accounting under Gaussian assumptions. Section~\ref{subsec:overshoot_main} establishes convergence to a Brownian reference under a separate set of stationary sampling and design assumptions.

\subsection{Oracle Gaussian Reference and the Deployed Gap}
\label{subsec:two_forms}

The oracle reference $C_m^*$ uses the same cumulative form as the deployed SPRT-z score $C_m$, with the baseline mean and arm variances replaced by their true values. Under the Gaussian and equal-maturation-fraction assumptions stated below, its directional form is Wald's Gaussian SPRT log-likelihood ratio for the mean difference \citep{wald1947}. Its two-sided form is the log of an equal-weight mixture of the two directional likelihood ratios. This provides a known-variance benchmark for the SPRT-z approximation motivated by Hajnal's sequential $t$-test \citep{hajnal1961}. The baseline mean $\mu$ fixes the target $\delta^*=r_{MDE}\mu$. Since $\mu$ and $\sigma_g^2$ are unknown in deployment, the reference is used for analysis. The gap $\Gamma_m=C_m-C_m^*$ can be computed in simulations with known population parameters.

Let look 0 contain the burn-in and let decisions occur at looks $1,\ldots,K$. Write $D_m=\bar x_{t,m}-\bar x_{c,m}$ for the cumulative arm-mean difference and $n_{g,m}$ for the matured count in arm $g$. For a stationary null with common mean $\mu$, set the fixed oracle target $\delta^*=r_{MDE}\mu$ and use the true arm variances:
\begin{equation}
\label{eq:oracle_reference}
\begin{aligned}
I_m^*&=\left(\frac{\sigma_t^2}{n_{t,m}}+\frac{\sigma_c^2}{n_{c,m}}\right)^{-1},\\
\psi_m^*&=\delta^*\sqrt{I_m^*},\qquad z_m^*=D_m\sqrt{I_m^*},\\
C_m^{*\pm}&=\pm\psi_m^*z_m^*-\tfrac12(\psi_m^*)^2,\\
C_m^{*(2)}&=\log\cosh(\psi_m^*z_m^*)-\tfrac12(\psi_m^*)^2,\\
M_m^{*\pm}&=\exp(C_m^{*\pm}),\qquad
M_m^{*(2)}=\tfrac12(M_m^{*+}+M_m^{*-})=\exp(C_m^{*(2)}).
\end{aligned}
\end{equation}
The target and variances are fixed before the burn-in is observed. The reference starts at $M_{-1}^{*\pm}=1$ and scores the burn-in as its first factor; there is no decision at look 0. In the two-sided reference, the directional sign is selected once for the whole path. The martingale conditions for $M_m^*$ are stated below.

The deployed score uses $\psi_m=r_{MDE}\bar x_{c,m-1}\sqrt{I_m}$ and $z_m=D_m\sqrt{J_m}$, where $I_m$ uses lagged sample variances and $J_m$ current sample variances; both are defined in Appendix~\ref{app:anatomy_proof}. Its directional and two-sided forms are
\begin{equation}
\label{eq:deployed_score}
\begin{aligned}
C_m^\pm&=\pm\psi_mz_m-\tfrac12\psi_m^2,\\
C_m^{(2)}&=\log\cosh(\psi_mz_m)-\tfrac12\psi_m^2.
\end{aligned}
\end{equation}
For valid denominators, the directional gap is the difference between the two scores:
\begin{equation}
\label{eq:gap_bridge}
\begin{aligned}
\Gamma_m^\pm&=C_m^\pm-C_m^{*\pm}
=\pm(\psi_mz_m-\psi_m^*z_m^*)
-\tfrac12\bigl(\psi_m^2-(\psi_m^*)^2\bigr).
\end{aligned}
\end{equation}
The gap identity holds for any outcome distribution when the variance denominators are valid. Appendix~\ref{app:anatomy_proof} separates $\Gamma_m^\pm$ into signed target and scale components and derives the two-sided gap $\Gamma_m^{(2)}$ from the directional gaps.

For the exact martingale result, newly revealed observations must be conditionally independent Gaussian variables within and between arms, with fixed positive arm variances and a common mean that may be predictable. Counts are fixed independently of outcomes and the maturation fractions must agree in the two arms at every look:
\begin{equation}
\label{eq:equal_fractions}
\frac{\Delta n_{t,k}}{n_{t,k}}=
\frac{\Delta n_{c,k}}{n_{c,k}},\qquad k=1,\ldots,K.
\end{equation}
Random assignment does not make realised fractions exactly equal. Under these conditions the directional references and their fixed mixture are positive martingales from before burn-in (Lemma~\ref{lem:oracle} in Appendix~\ref{app:comp_setup}). With unequal fractions, the same cumulative formula remains a useful comparator but need not be a martingale.

\subsection{Exact Rejection Accounting and Overshoot}
\label{subsec:compensator_main}

We express the deployed null rejection probability $P_0(E)$, where $E=\{C_T\ge A\}$, in terms of the score gap $\Gamma=C_T-C_T^*$ and efficacy overshoot $R=C_T-A$ on $E$.

Here $T$ is the first look at which the deployed score crosses either Wald boundary $A$ or $B$, defined in Section~\ref{subsec:boundaries}, or the final look $K$ if neither boundary is crossed. A crossing at $K$ takes precedence over truncation. The deployed score and oracle reference use the same directional or two-sided form.

Under the Gaussian conditions above, weighting each possible experiment path by the oracle likelihood ratio $M_K^*$ defines a reweighted distribution $Q^*$. The weights average to one under the null distribution $P_0$, so the resulting probabilities sum to one. This change of measure is written $dQ^*/dP_0=M_K^*$. For $P_0(E)>0$, applying it at the bounded stopping look $T$ gives the identity
\begin{equation}
\label{eq:accounting}
P_0(E)=\frac{\alpha}{1-\beta}\,
Q^*(E)\,E_{Q^*}[e^{\Gamma-R}\mid E]
=\frac{\alpha}{1-\beta}\,
\frac{Q^*(E)}{E_{P_0}[e^{R-\Gamma}\mid E]}.
\end{equation}
Equation~\ref{eq:accounting} expresses the deployed Type I error $P_0(E)$ through the reweighted rejection probability $Q^*(E)$ and the gap-and-overshoot adjustment $E_{Q^*}[e^{\Gamma-R}\mid E]$. Under the stated Gaussian assumptions, $P_0(E)\le\alpha$ exactly when
\begin{equation}
\label{eq:oracle_control}
Q^*(E)E_{Q^*}[e^{\Gamma-R}\mid E]\le 1-\beta.
\end{equation}
At $(\alpha,\beta)=(0.05,0.20)$, this means that the product must be at most $0.80$ to achieve Type I error at or below $5\%$. Both quantities can be estimated in Gaussian simulations with known population parameters, providing a direct way to assess the deployed stopping rule. Futility and horizon truncation enter through the rejection event $E$; the factor $e^{\Gamma-R}$ accounts for the score gap and overshoot.

For comparison, applying Ville's inequality to the oracle reference alone gives a $6.25\%$ bound at the Wald boundary $A=\log16$. Equation~\ref{eq:oracle_control} gives the exact criterion for $5\%$ control of the deployed rule by incorporating its stopping decisions, score gap, and overshoot. Appendix~\ref{app:type_i_accounting} provides the derivation, the interpretation of $Q^*$, and the scope of this criterion.

\subsection{Fixed-Look Approximation for Non-Gaussian Outcomes}
\label{subsec:overshoot_main}

Section~\ref{subsec:compensator_main} gives exact rejection accounting for Gaussian observations. For skewed outcomes, we need to establish when the same reference approximates the deployed test. Classical Brownian approximations provide the connection by describing cumulative evidence on an information-time scale \citep{siegmund1985sequential, jennison1999group}.

We compare progressively larger experiments designed to detect progressively smaller effects. This allows the baseline and variance estimates to improve while keeping the standardized target on a comparable scale. Specifically, post-burn-in traffic $N$ and burn-in size $\nu_N$ grow, while the relative MDE shrinks as $r_{MDE,N}=c/\sqrt N$, for fixed $c>0$. The number of looks $K$, allocation proportions, and post-burn-in cohort proportions remain fixed. Appendix~\ref{app:local_proof} specifies the burn-in scaling.

Outcomes are stationary and i.i.d.\ within each arm. Under the null, the arms share a nonzero mean and have positive finite variances; their distribution shapes and variances may differ. Under this design sequence, the cumulative mean difference $D_{m,N}$, weighted by its oracle information $I_{m,N}^*$, satisfies
\[
\frac{I_{m,N}^*D_{m,N}}{\sqrt{I_{K,N}^*}}
\;\Rightarrow\; W_{t_m},
\qquad
 t_m=\lim_{N\to\infty}\frac{I_{m,N}^*}{I_{K,N}^*},
\]
jointly across the monitored looks. Here $W_t$ is standard Brownian motion and $t_m$ is the limiting fraction of total information available at look $m$. Substituting this limit into the score formulas in Section~\ref{subsec:two_forms} gives the directional and two-sided Brownian reference scores on the same monitoring grid.

As the baseline and variance estimates become accurate, the deployed--oracle gap vanishes. Together with the joint Brownian limit, this yields
\[
\max_{1\le m\le K}|\Gamma_{m,N}|
\xrightarrow{p}0,
\qquad
P_0(E_N)\longrightarrow p_\infty,
\]
where $E_N$ is the deployed rejection event and $p_\infty$ is the limiting null rejection probability as $N\to\infty$ (Theorem~\ref{thm:local_oracle}, Appendix~\ref{app:local_proof}). Thus, the deployed Type I error approaches that of the Brownian rule with the same boundaries and final-look convention.

Applying the accounting identity from Section~\ref{subsec:compensator_main} directly to that Brownian rule, whose score gap is zero, gives
\begin{equation}
\label{eq:brownian_accounting}
p_\infty=\frac{\alpha}{1-\beta}q_\infty u_\infty.
\end{equation}
Here $q_\infty=Q_\infty(E_\infty)$ is the reweighted rejection probability, and $u_\infty=E_{Q_\infty}[e^{-R_\infty}\mid E_\infty]$ is the mean overshoot correction among rejected paths. Discrete crossings give $0<u_\infty<1$. If calibration achieves $q_\infty=1-\beta$, then $p_\infty=\alpha u_\infty<\alpha$. More generally, nominal control holds exactly when $q_\infty u_\infty\le1-\beta$. Appendix~\ref{app:comp_audit} evaluates these quantities for the specified Brownian rule.

The stationary lognormal-mixture Negative Binomial model satisfies the theorem's moment assumptions, supporting this comparison along the specified design sequence. Sections~\ref{sec:empirical} and~\ref{sec:bias} assess finite-sample performance through simulation. Appendix~\ref{app:local_proof} gives the full assumptions, convergence-rate results, and limitations.

\section{Empirical Validation of SPRT-z}
\label{sec:empirical}

We compare SPRT-z with SFHC against FHT under a skewed, overdispersed outcome model with staggered enrolment and fixed observation windows. The simulations assess whether horizon calibration preserves power and how early stopping changes average sample use across effect sizes and daily batch sizes.

\subsection{Simulation Design}\label{sec:simframework}
We simulate 2,500 replications per effect and daily batch size, with a user pool of $1.1N_{max}$ and equal allocation between arms. All sample sizes are totals across both arms. The data-generating process (DGP) has four features:
\begin{enumerate}
    \item \textbf{Right skew:} unit action rates are Log-Normal with $\mu_{log}=\ln(3)$ and $\sigma_{log}=1.1$.
    \item \textbf{Overdispersion:} realized values are Negative Binomial with the unit action rate and dispersion $\theta=1$.
    \item \textbf{Staggered entry:} weekend enrolment is one tenth of the weekday rate.
    \item \textbf{Cohort maturation:} each unit enters the test only after its seven-day observation window closes, producing disjoint cohorts; the simulation draws their outcomes independently.
\end{enumerate}

We use two-sided tests with inputs $\alpha=0.05$ and $\beta=0.20$, a relative design MDE of $\delta=0.10$, and daily batch sizes $n_{daily}\in\{100,500,2500\}$. We calculate $N_{FHT}$ from the empirical baseline variance and use SFHC to choose $N_{max}$ to approximately match FHT power, allowing for futility stopping and discrete monitoring. Both tests use the same simulated experiments.

We repeat the study with a bounded Bernoulli outcome, baseline $p_0=0.30$, relative lifts $\delta\in\{0,0.05,0.10,0.15\}$, and the same staggered entry, cohort structure, and $n_{daily}\in\{100,500,2500\}$. Appendix~\ref{app:simulation_details} reports the results.

\begin{figure}[htbp]
    \centering
    \includegraphics[width=\linewidth]{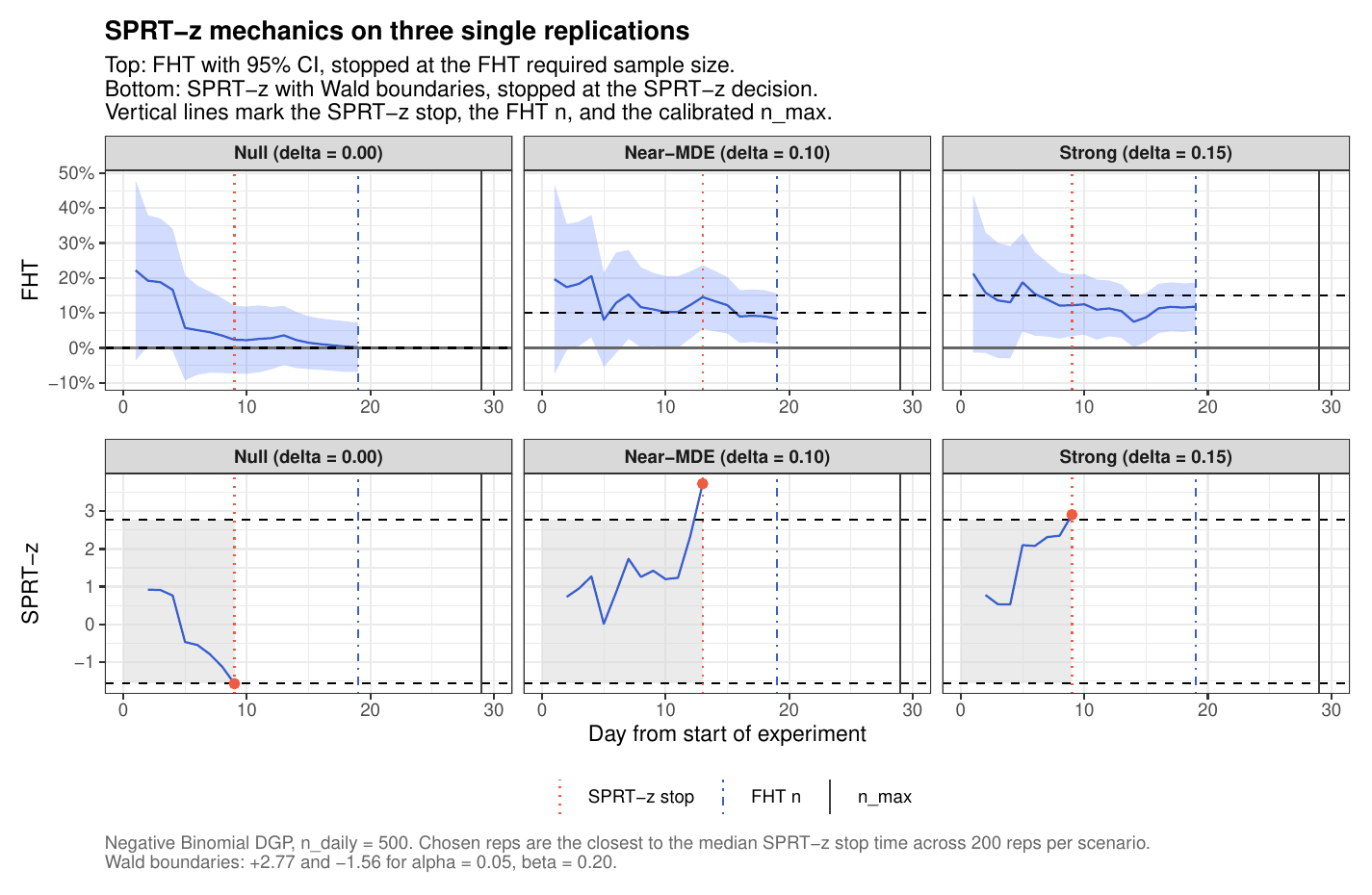}
    \caption{FHT and SPRT-z on three Negative Binomial sample paths at $n_{daily}=500$, with two-sided inputs $\alpha=0.05$ and $\beta=0.20$. Each column shows the replication closest to the median SPRT-z stopping time among 200 separately simulated replications for that effect. The upper panels show running relative-effect estimates and pointwise 95\% Wald intervals for illustration; FHT makes its decision only at $N_{FHT}$. The lower panels show the SPRT-z log-likelihood ratio up to its first boundary crossing. Vertical lines mark the SPRT-z stop, $N_{FHT}$, and $N_{max}$.}
    \label{fig:trajectory}
\end{figure}

Figure~\ref{fig:trajectory} illustrates how the two tests use the same Negative Binomial data at $n_{daily}=500$. In the selected null replication, SPRT-z crosses the lower boundary for futility on day 9. At the design MDE ($\delta=0.10$), it crosses the upper boundary on day 13; at $\delta=0.15$, it does so on day 9. FHT reaches its planned sample size on day 19 in all three examples, while the calibrated SPRT-z horizon extends to day 29. These paths illustrate the stopping rule; Section~\ref{csd:simresult} evaluates error rates, power, and average sample use across the full simulation grid.

\subsection{Error Rates, Power, and Sample Use}
\label{csd:simresult}
Table~\ref{tab:validation} shows that a larger maximum sample size can accompany a smaller average sample size. Although SFHC sets $N_{max}>N_{FHT}$, early stopping reduces average sample use in all but one evaluated setting. Savings are largest at the null and above the design MDE, ranging from about 19\% to 46\% across those settings.

Larger daily batches reduce the opportunities to realise early-stopping savings: a boundary crossing that finer monitoring would detect partway through a batch can only be acted on at the next scheduled analysis. The additional observations can consume much or all of the potential saving relative to FHT. Intermediate effects may also remain undecided beyond the fixed-horizon sample size because SFHC permits a longer maximum horizon. At half the MDE under the coarsest monitoring evaluated here, average sample use slightly exceeds FHT.

At the design MDE, SFHC brings SPRT-z power within three percentage points of FHT in all three settings. This agreement is approximate: paired comparisons identify small differences beyond Monte Carlo uncertainty at the two smaller batch sizes, while the difference at the largest batch size is indistinguishable from simulation noise. Matching FHT power also does not ensure the nominal 80\% target. At $n_{daily}=100$, for example, FHT attains 74.4\% power and SPRT-z 77.1\%. Under this heavily skewed DGP, the normal-theory sample-size calculation with an estimated baseline variance does not consistently deliver the intended FHT power; the shortfall is smaller at larger daily batch sizes. Calibration against that benchmark therefore requires checking realised power for the outcome model and monitoring schedule.

Null rejection rates are below 5\% at every evaluated batch size and decrease as batches grow. This pattern is compatible with discrete-look overshoot and futility stopping, although the comparison does not isolate their effects. At the smallest batch size, roughly fifty units mature per arm per day, so the strongly skewed outcomes also limit the Gaussian approximation. Error control in that setting is supported by simulation rather than by an exact Gaussian theorem (Appendix~\ref{app:comp_audit}). All null cells here use identically distributed arms. These results do not establish control for every equal-mean null: Appendix~\ref{app:comp_attack} reports excess rejection in some settings with unequal distributional shapes across arms.

\begin{table}[H]
    \caption{Comparative Performance of FHT vs. SPRT-z (2,500 Simulations per Cell, Stationary Baseline). The Rejection Rate column reports the empirical Type I error at $\delta = 0.00$ (target $\alpha = 0.05$) and the statistical power at $\delta > 0$ (nominal target $1 - \beta = 0.80$ at the MDE). Each rate is shown with its 95\% Monte Carlo half-width, $\pm 1.96\sqrt{\hat{p}(1-\hat{p})/2500}$. The power shortfall at the MDE is most apparent at the two smaller daily batch sizes; see the text for discussion.}
    \centering
    \small 
    \begin{tabular}{lrcrrrr}
        \toprule
        & & \multicolumn{2}{c}{\textbf{Rejection Rate}} & \multicolumn{2}{c}{\textbf{Average Sample Size}} & \\
        \cmidrule(lr){3-4} \cmidrule(lr){5-6}
        \textbf{Daily Batch} & \textbf{True Effect ($\delta$)} & \textbf{FHT} & \textbf{SPRT-z} & \textbf{FHT} & \textbf{SPRT-z} & \textbf{\% Reduction} \\
        \midrule
        \textbf{$n_{daily} = 100$}
        & 0.00 (Null) & 0.050 {\scriptsize$\pm$.009} & 0.041 {\scriptsize$\pm$.008} & 8,379 & 5,390 & 35.7\% \\
        & 0.05        & 0.263 {\scriptsize$\pm$.017} & 0.241 {\scriptsize$\pm$.017} & 8,377 & 6,707 & 19.9\% \\
        & 0.10 (MDE)  & 0.744 {\scriptsize$\pm$.017} & 0.771 {\scriptsize$\pm$.016} & 8,377 & 6,408 & 23.5\% \\
        & 0.15        & 0.972 {\scriptsize$\pm$.006} & 0.972 {\scriptsize$\pm$.006} & 8,379 & 4,502 & 46.3\% \\
        \midrule
        \textbf{$n_{daily} = 500$}
        & 0.00 (Null) & 0.049 {\scriptsize$\pm$.008} & 0.025 {\scriptsize$\pm$.006} & 9,099 & 5,969 & 34.4\% \\
        & 0.05        & 0.268 {\scriptsize$\pm$.017} & 0.233 {\scriptsize$\pm$.017} & 9,097 & 7,610 & 16.3\% \\
        & 0.10 (MDE)  & 0.776 {\scriptsize$\pm$.016} & 0.756 {\scriptsize$\pm$.017} & 9,097 & 7,166 & 21.2\% \\
        & 0.15        & 0.981 {\scriptsize$\pm$.005} & 0.976 {\scriptsize$\pm$.006} & 9,097 & 4,957 & 45.5\% \\
        \midrule
        \textbf{$n_{daily} = 2500$}
        & 0.00 (Null) & 0.050 {\scriptsize$\pm$.009} & 0.019 {\scriptsize$\pm$.005} & 9,612 & 7,792 & 18.9\% \\
        & 0.05        & 0.300 {\scriptsize$\pm$.018} & 0.228 {\scriptsize$\pm$.016} & 9,609 & 9,904 & -3.1\% \\
        & 0.10 (MDE)  & 0.794 {\scriptsize$\pm$.016} & 0.789 {\scriptsize$\pm$.016} & 9,611 & 9,280 & 3.4\% \\
        & 0.15        & 0.984 {\scriptsize$\pm$.005} & 0.988 {\scriptsize$\pm$.004} & 9,614 & 6,997 & 27.2\% \\
        \bottomrule
    \end{tabular}
    \label{tab:validation}
\end{table}

\raggedbottom
\section{Bias Correction and Confidence Intervals}
\label{sec:bias}

Stopping when results look promising can exaggerate the estimated effect, while futility stopping can pull estimates toward zero. Ordinary fixed-horizon confidence intervals also lose their nominal coverage when applied at a data-dependent stopping time. This section adjusts effect estimates and intervals for the SPRT-z stopping rule, including experiments that reach the maximum horizon.\footnote{The adjustment addresses bias from sequential stopping. Selective reporting across metrics or segments \citep{dengPostSelection2021} and selection among experiments \citep{leeShen2018} are separate sources of bias.}

Stagewise ordering ranks outcomes using the terminal decision, stopping time, and final statistic. Classical group sequential constructions yield exact confidence intervals and a median unbiased estimator (MUE) under their assumed sampling model and stopping rule \citep{tsiatis1984exact, jennison1999group}. We use a six-region ordering for SPRT-z: efficacy, futility, and horizon-truncated outcomes, each split by the sign of the final statistic. The ordering runs from negative to positive evidence and is defined in Appendix~\ref{app:mue}.

The calculation requires probabilities of stopping outcomes, traditionally evaluated by numerical integration \citep{armitage1969significance}. We estimate these probabilities by Monte Carlo simulation, using the Brownian approximation introduced for SFHC in Section~\ref{sec:horizon} \citep{siegmund1985sequential, lan1983discrete}. For each candidate true effect, we simulate Brownian paths with the corresponding drift, check the SPRT-z boundaries at discrete monitoring steps, and record each path's stopping outcome. The proportion ranked at least as positive as the observed outcome gives a one-sided tail probability for that candidate effect. Inference for the observed data is approximate because the Brownian model replaces the empirical process.

The MUE and confidence interval invert this same probability calculation. The MUE is the effect at which the tail probability is 0.50; the lower and upper 95\% interval endpoints use 0.025 and 0.975. We find these effects by bisection and keep comparisons stable by reusing the same random draws across candidates, a technique known as Common Random Numbers (CRN) \citep{glasserman2004monte}. Algorithms~\ref{alg:pvalue} and~\ref{alg:bisection} in Appendix~\ref{app:mue} give the implementation details.

\subsection{Point Estimation and Bias Reduction}

\begin{figure}[h]
    \centering
    \includegraphics[width=\linewidth]{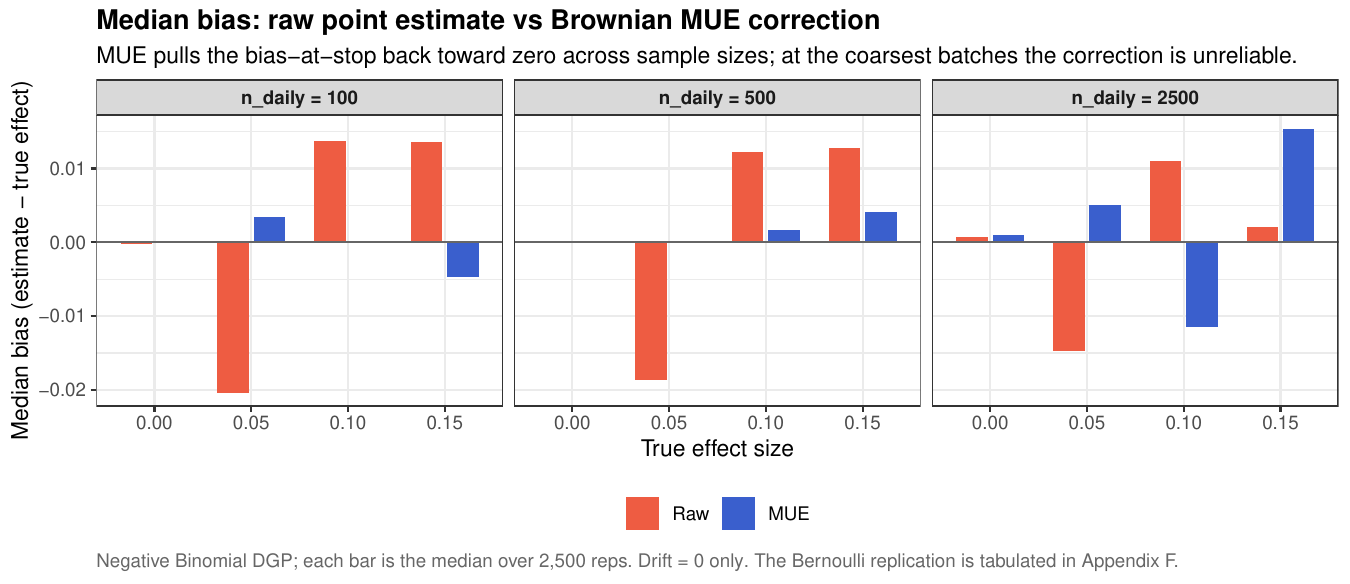}
    \caption{Median bias of the SPRT-z stopped estimate under the Negative Binomial DGP, raw versus Brownian MUE, across $n_{daily} \in \{100, 500, 2500\}$ and $\delta \in \{0, 0.05, 0.10, 0.15\}$. Each bar is the median over 2{,}500 replications under the stationary baseline; the Bernoulli replication is tabulated in Appendix~\ref{app:simulation_details}.}
    \label{fig:bias_reduction}
\end{figure}

Figure~\ref{fig:bias_reduction} compares median bias before and after MUE adjustment in the simulations of Section~\ref{sec:simframework}. Information-time resolution describes how much of the experiment's total information arrives between checks: smaller batches give finer steps, and larger batches give coarser steps. All three settings use daily checks, but larger daily batches leave fewer looks before the horizon.

With smaller daily batches, the MUE substantially reduces median bias. At $\delta=0.10$ and $n_{daily}=100$, bias falls from 0.0138 to 0.0003.

At the underpowered effect $\delta=0.05$, futility stopping pulls raw estimates toward zero. The MUE reduces this negative bias by 83\% and 99\% at $n_{daily}=100$ and $500$, respectively.

At $n_{daily}=2500$, the adjustment can increase bias. At $\delta=0.10$, the MUE changes bias from $+0.011$ to $-0.012$; at $\delta=0.15$, it changes $+0.002$ to $+0.015$. The Bernoulli DGP shows a similar overcorrection with large batches (Appendix~\ref{app:simulation_details}). These results identify a limitation when tests have few looks, but do not isolate its cause: the Brownian simulation already checks boundaries at discrete steps and allows overshoot.

\begin{figure}[h]
    \centering
    \includegraphics[width=\linewidth]{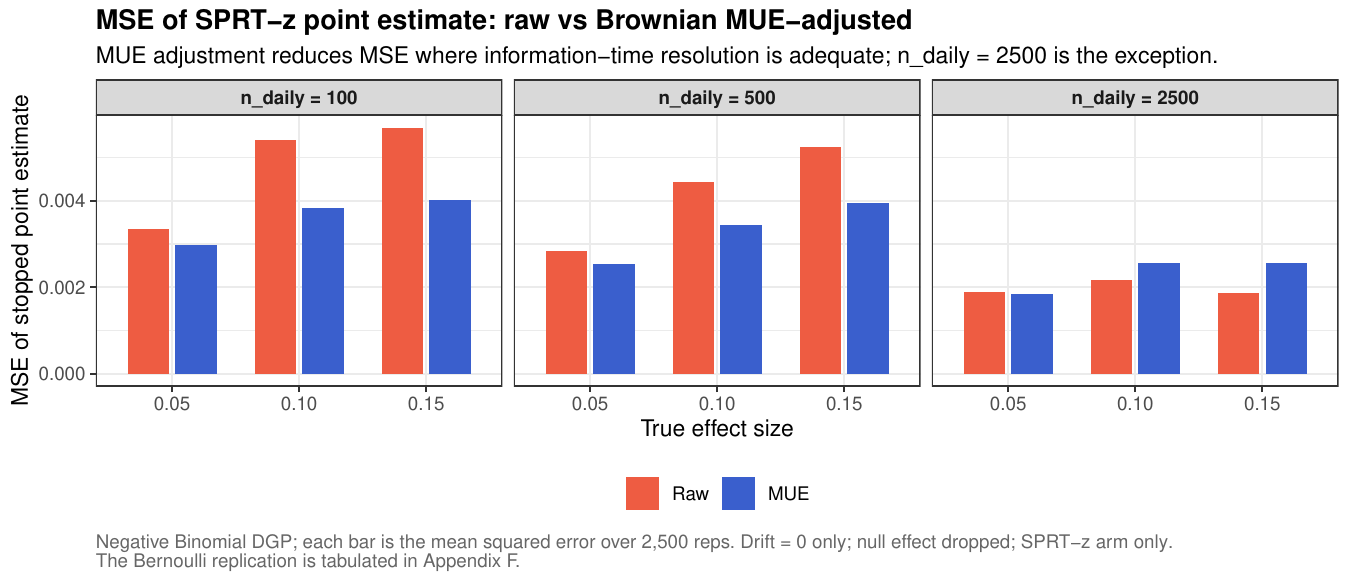}
     \caption{Mean Squared Error of the raw SPRT-z and MUE-adjusted SPRT-z stopped estimate under the Negative Binomial DGP, across the non-null cells of the evaluation grid in Figure~\ref{fig:bias_reduction}. The MUE-adjusted MSE sits below the raw SPRT-z MSE across most cells; the exception is the large-batch regime at $n_{daily} = 2500$, mirroring the bias miscorrection documented in Figure~\ref{fig:bias_reduction}.}
    \label{fig:mse_summary}
\end{figure}

Figure~\ref{fig:mse_summary} shows that the MUE also lowers Mean Squared Error (MSE) in the settings with smaller daily batches.

At $\delta=0.10$ and $n_{daily}=500$, adjusted MSE ($3.43\times10^{-3}$) remains above FHT MSE ($1.49\times10^{-3}$). Early stopping uses less data and therefore has greater variance; the MUE corrects bias, not variance.

\subsection{Confidence Interval Coverage}
The interval contains the candidate effects whose simulated tail probabilities lie between 0.025 and 0.975. Its coverage is the proportion of repeated experiments in which this interval contains the true effect. For each effect and daily batch size in Section~\ref{sec:simframework}, we run 2,500 experiments to efficacy, futility, or the maximum horizon and construct one interval at each experiment's stopping look. The 95\% target applies across these outcomes combined (unconditional coverage). These are intervals at the stopping look; they do not provide simultaneous coverage at every interim look, as a confidence sequence would.

Table~\ref{tab:ci_coverage} reports the resulting coverage under the stationary baseline. The Monte Carlo half-widths describe uncertainty in the estimated coverage rates.

\begin{table}[H]
    \caption{Empirical Unconditional Coverage of 95\% Brownian MUE Confidence Intervals (Stationary Baseline, 2,500 Simulations per Cell). Each rate is shown with its 95\% Monte Carlo half-width in percentage points.}
    \centering
    \small 
    \begin{tabular}{lrr}
        \toprule
        \textbf{Daily Batch} & \textbf{True Effect ($\delta$)} & \textbf{CI Coverage} \\
        \midrule
        \textbf{$n_{daily} = 100$}
        & 0.00 (Null) & 94.7\% {\scriptsize$\pm$0.9} \\
        & 0.05        & 95.6\% {\scriptsize$\pm$0.8} \\
        & 0.10 (MDE)  & 94.7\% {\scriptsize$\pm$0.9} \\
        & 0.15        & 96.2\% {\scriptsize$\pm$0.7} \\
        \midrule
        \textbf{$n_{daily} = 500$}
        & 0.00 (Null) & 95.7\% {\scriptsize$\pm$0.8} \\
        & 0.05        & 94.0\% {\scriptsize$\pm$0.9} \\
        & 0.10 (MDE)  & 94.3\% {\scriptsize$\pm$0.9} \\
        & 0.15        & 96.2\% {\scriptsize$\pm$0.7} \\
        \midrule
        \textbf{$n_{daily} = 2500$}
        & 0.00 (Null) & 96.0\% {\scriptsize$\pm$0.8} \\
        & 0.05        & 94.9\% {\scriptsize$\pm$0.9} \\
        & 0.10 (MDE)  & 98.0\% {\scriptsize$\pm$0.5} \\
        & 0.15        & 94.3\% {\scriptsize$\pm$0.9} \\
        \bottomrule
    \end{tabular}
    \label{tab:ci_coverage}
\end{table}

Table~\ref{tab:ci_coverage} reports coverage between 94.0\% and 98.0\% across the twelve settings. Seven cells are consistent with 95\% within Monte Carlo uncertainty; four show conservative coverage and one shows modest undercoverage. These results support approximate 95\% coverage in the evaluated settings, with a maximum observed shortfall of one percentage point.

\raggedbottom
\section{SPRT-z in Production at Atlassian}
\label{sec:production}

The deployment provides a comparison with the fixed-horizon plans that the same experiments would otherwise have used.

\subsection{Where SPRT-z sits in the platform}
\label{subsec:deployment}

Experiment owners select SPRT-z when configuring an experiment in Atlassian's platform; this section therefore covers experiments whose owners chose it over FHT. Teams set the relative MDE through the same pre-experiment power analysis used for FHT. The running control arm supplies the baseline that converts it to absolute $\delta_{MDE}$, following Section~\ref{sec:prequential}.

\subsection{Operating characteristics against the fixed-horizon plan}
\label{subsec:postlaunch}

For each of $46$ comparable experiments, we calculate the sample at its SPRT-z stopping rule and the sample required by its fixed-horizon plan from recorded daily arm series. The fixed-horizon requirement uses that experiment's declared target and measured control-arm dispersion. Every number in this section uses these $46$ experiments.

\textbf{Stopping mix.} Of 46 experiments, 38 crossed the futility boundary and eight crossed efficacy. Consistent with the 70--90\% feature failure rates reported elsewhere \citep{kohavi2020trustworthy}, futility stopping produced most decisions.

\begin{figure}[H]
    \centering
    \includegraphics[width=0.92\linewidth]{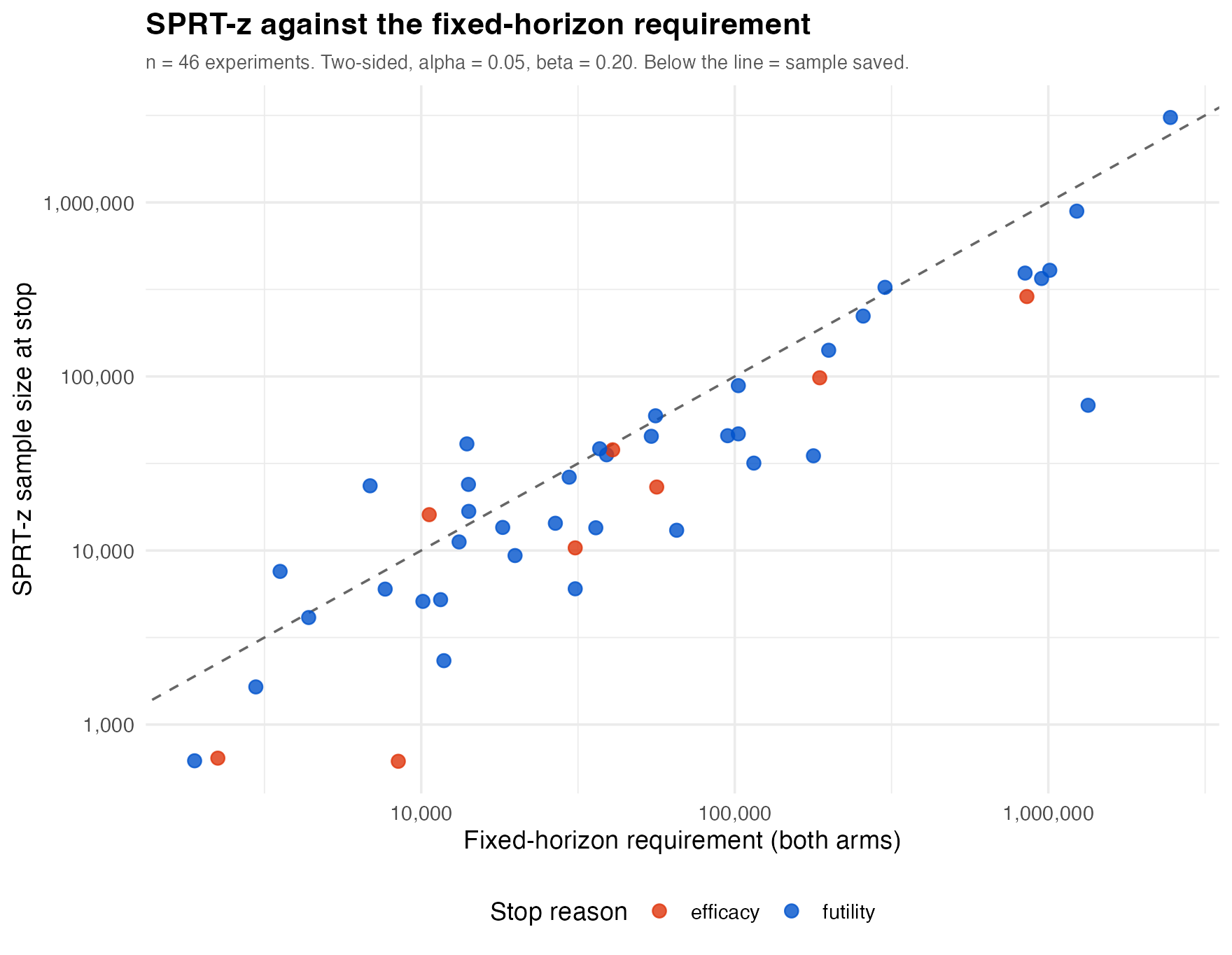}
    \caption{Realised SPRT-z stop size against the fixed-horizon sample requirement, total across both arms, on log-log axes, for the same 46 experiments as Table~\ref{tab:production_savings}. The dashed line marks equality, so points below it saved sample relative to the fixed-horizon plan. Colour denotes the stopping rule that fired. Points above the line are runs on which the statistic lingered in the continuation region and spent more than the fixed-horizon plan would have, which Table~\ref{tab:validation} reproduces in simulation at half the design MDE.}
    \label{fig:production_savings}
\end{figure}

\textbf{Sample size recovered.} The median experiment used $54.5\%$ of its fixed-horizon requirement, saving $45.5\%$ (IQR $7.6\%$ to $62.3\%$; Table~\ref{tab:production_savings}). Of $46$ experiments, $36$ saved sample. The worst used $3.4$ times the requirement and the best saved $94.9\%$. Median savings were $62.6\%$ for efficacy crossings and $28.4\%$ for futility crossings, consistent with clear effects reaching the boundary fastest (Table~\ref{tab:validation}). The $10$ runs above the equality line in Figure~\ref{fig:production_savings} show the cost of longer continuation on individual experiments.

\begin{table}[ht]
\caption{Sample size recovered against the fixed-horizon requirement, on the 46 experiments the comparison covers. Savings are in total sample across both arms.}
\label{tab:production_savings}
\centering
\small
\setlength{\tabcolsep}{5pt}
\begin{tabular}{lrrr}
\toprule
& $n$ & Median saving & Range \\
\midrule
Efficacy crossings & 8  & 62.6\% & $-51.5\%$ to 92.7\% \\
Futility crossings & 38 & 28.4\% & $-242.6\%$ to 94.9\% \\
\midrule
All experiments     & 46 & 45.5\% & IQR 7.6\% to 62.3\% \\
\bottomrule
\end{tabular}
\end{table}

Because live true effects are unobserved, these data measure decision cost, not production error rates.

\section{Conclusion}
\label{sec:conclusion}
SPRT-z adapts Wald's two-boundary test to online experiments in which teams need early efficacy and futility decisions against a business-relevant design MDE. The normal approximation reduces the cost of evaluating the statistic and supports Brownian calculations for planning and estimation. SFHC sets a maximum sample size to approximately recover the fixed-horizon anchor's power, allowing for futility stopping and batch monitoring. Inversion of simulated stopping probabilities then provides effect estimates and intervals adjusted for the stopping rule.

The simulations show how a larger maximum sample size can coexist with lower average sample use. Early stopping saved sample in most evaluated settings, while power generally remained close to FHT except under the coarsest monitoring. Intermediate effects could require longer continuation and sometimes use more sample than FHT. The production results support the practical value of futility stopping, which accounted for $38$ of $46$ decisions. Across those experiments, the median sample saving against fixed-horizon plans was $45.5\%$. These observations measure decision cost; the unknown true effects prevent an assessment of production error rates.

The validity analysis separates the deployed cumulative statistic from a Gaussian oracle with a fixed target and true variances. It gives exact rejection accounting under the oracle assumptions and a fixed-look approximation for stationary finite-variance outcomes (Section~\ref{sec:validity}). Neither result gives finite-sample control for every equal-mean null. Power also depends on calibration, since SFHC's fixed-horizon anchor may itself miss the nominal target. Both error rates therefore require validation for the intended outcome distribution, monitoring cadence, and horizon (Appendix~\ref{app:comp_attack}).

Estimation after stopping requires a separate assessment. The Brownian MUE reduced bias and its intervals generally achieved approximately nominal coverage in the evaluated settings. With few looks, however, the adjustment could increase bias and coverage could depart materially from nominal. The simulations identify these failures without isolating their cause. The intervals target unconditional coverage at the stopping look; they do not provide simultaneous coverage throughout monitoring.

The method is developed for matured observations collected over fixed windows, under the independence assumptions in Section~\ref{subsec:fixedobswindow}. Extending it to open-window longitudinal panels requires accounting for dependence within units. For estimation, a useful next step is to evaluate inversion that reproduces the empirical variance updates and prequential targets more closely, particularly when experiments stop after only a few looks.

\section{Acknowledgements}

We thank Marianne Menictas (Adobe), Adam Gustafson (Microsoft), Dylan Lewis, and colleagues at Atlassian for feedback and support. The views expressed are those of the authors and not necessarily those of the reviewers or their organisations.

\appendix
\raggedbottom

\section{Type I Error Analysis and Proofs}
\label{app:compensator}

This appendix proves the oracle construction, deployed-gap decomposition, exact rejection accounting, and fixed-look local limit used in Section~\ref{sec:validity}. The exact Gaussian and stationary asymptotic statements have different assumptions.

\subsection{Oracle Construction and Initialisation}
\label{app:comp_setup}

Look 0 contains the burn-in, with $n_{g,0}\ge2$ in each arm. Let $Y_k=\bar y_{t,k}-\bar y_{c,k}$ be the difference for the newly matured cohort at look $k\ge1$, and put $Y_0=D_0$. Counts and cohort membership are determined independently of outcomes. The filtration $\mathcal F_{-1}$ contains the fixed design but no experiment outcomes, while $\mathcal F_m$ contains outcomes matured through look $m$. For the exact Gaussian result, conditional on $\mathcal F_{k-1}$, newly revealed observations in the two arms are independent Gaussian variables with a common predictable mean $\mu_k$ and fixed positive arm variances $\sigma_t^2,\sigma_c^2$. This also applies to the burn-in relative to $\mathcal F_{-1}$.

\begin{lemma}[Oracle Gaussian martingale]
\label{lem:oracle}
If the realised counts satisfy Equation~\ref{eq:equal_fractions}, each directional process $M_m^{*\pm}$ in Equation~\ref{eq:oracle_reference} and their fixed 50/50 mixture $M_m^{*(2)}$ are positive martingales starting at $M_{-1}^*=1$. The target $\delta^*$ and true variances are fixed before observing burn-in.
\end{lemma}

\begin{proof}
Set $H_0=I_0^*$ and, for $k\ge1$, set
\[
H_k=\left(\frac{\sigma_t^2}{\Delta n_{t,k}}+
\frac{\sigma_c^2}{\Delta n_{c,k}}\right)^{-1}.
\]
With $q_k=\Delta n_{g,k}/n_{g,k}$ common to the two arms, $I_{k-1}^*=(1-q_k)I_k^*$, $H_k=q_kI_k^*$, and $D_k=(1-q_k)D_{k-1}+q_kY_k$. Therefore $I_m^*=\sum_{k=0}^mH_k$ and $I_m^*D_m=\sum_{k=0}^mH_kY_k$. Each reference factor is
\[
\exp\{\pm\delta^*H_kY_k-\tfrac12\delta^{*2}H_k\}.
\]
The common mean cancels from $Y_k$, whose conditional law is $N(0,H_k^{-1})$. The Gaussian moment-generating identity makes every factor's conditional mean one. This includes $k=0$, since the target and true variances precede burn-in. Multiplication proves the directional martingales, and linearity proves the mixture. Bounded optional sampling and Ville's inequality then apply. The mixture averages whole-path directional products; multiplying separately remixed cohort factors would define a different process.
\end{proof}

The lemma also permits an outcome-independent random schedule conditional on realised counts that satisfy Equation~\ref{eq:equal_fractions} and preserve the conditional Gaussian model. Ordinary random allocation does not ensure this equality. With unequal fractions the cumulative expression in Equation~\ref{eq:oracle_reference} generally has no martingale property; the empirical replay in Appendix~\ref{app:comp_audit} uses it only as a comparator. The predictable common-mean extension in the lemma does not give consistency of estimated means or variances under drift.

\subsection{Target and Scale Components of the Gap}
\label{app:anatomy_proof}

For $m\ge1$, the deployed lagged and current inverse squared standard errors are
\begin{equation}
\label{eq:score_scales}
I_m=\left(\frac{s_{t,m-1}^2}{n_{t,m}}+
\frac{s_{c,m-1}^2}{n_{c,m}}\right)^{-1},\qquad
J_m=\left(\frac{s_{t,m}^2}{n_{t,m}}+
\frac{s_{c,m}^2}{n_{c,m}}\right)^{-1}.
\end{equation}
The identities require finite positive combined denominators; an individual arm variance may be zero. To split Equation~\ref{eq:gap_bridge} into target and scale components, use the intermediate score $\bar C_m^\pm=\pm\delta^*\sqrt{I_m}\,z_m-\delta^{*2}I_m/2$. It combines the fixed oracle target with the deployed scales. Subtracting first $\bar C_m^\pm$ and then $C_m^{*\pm}$ gives
\[
C_m^\pm-\bar C_m^\pm
=\pm(\psi_m-\delta^*\sqrt{I_m})z_m
-\tfrac12(\psi_m^2-\delta^{*2}I_m),\qquad
\bar C_m^\pm-C_m^{*\pm}
=\pm(\delta^*\sqrt{I_m}\,z_m-\psi_m^*z_m^*)
-\tfrac12\bigl(\delta^{*2}I_m-(\psi_m^*)^2\bigr).
\]
The first difference changes the target at the deployed scales; the second changes the scales at the fixed target. They sum to $\Gamma_m^\pm$. This algebra uses no distributional assumption and holds with unequal realised counts. Both terms are signed. The exact oracle depends on the null mean and true variances, so it is not available as a production statistic.

For the two-sided score, define
$\omega_m=M_m^{*+}/(M_m^{*+}+M_m^{*-})$. The mixture gap is
\begin{equation}
\label{eq:two_gap}
\Gamma_m^{(2)}
=\log\{\omega_me^{\Gamma_m^+}+(1-\omega_m)e^{\Gamma_m^-}\}.
\end{equation}
It lies between the two directional gaps. An additive target and scale decomposition is obtained by using the two-sided intermediate score
$\bar C_m^{(2)}=\log\cosh(\delta^*\sqrt{I_m}\,z_m)-\delta^{*2}I_m/2$ and subtracting it in turn. Its components are not averages of the directional components.

\subsection{Change of Measure and Stopping Accounting}
\label{app:type_i_accounting}

Use the deployed stopping time and rejection event from Section~\ref{subsec:compensator_main}. Under Lemma~\ref{lem:oracle}, $M_K^*$ has mean one and defines $Q^*$ on $\mathcal F_K$. Since $T$ is bounded, $M_T^*$ is its density on $\mathcal F_T$. On $E$, $M_T^*=\exp(A+R-\Gamma)$. Integrating $\mathbbm{1}_E/M_T^*$ under $Q^*$ gives
\[
P_0(E)=e^{-A}Q^*(E)E_{Q^*}[e^{\Gamma-R}\mid E].
\]
Integrating $\mathbbm{1}_EM_T^*$ under $P_0$ gives
$Q^*(E)=e^AP_0(E)E_{P_0}[e^{R-\Gamma}\mid E]$.
These are the two forms of Equation~\ref{eq:accounting}. The stopped density incorporates paths ending at futility or the horizon. Since the Gaussian reference is a martingale rather than merely a supermartingale,
\[
1=e^AP_0(E)E_{P_0}[e^{R-\Gamma}\mid E]
  +E_{P_0}[M_T^*\mathbbm{1}_{E^c}].
\]
Dropping the last term gives the weaker bound
\begin{equation}
\label{eq:overshoot_bound}
P_0(E)\le
\frac{e^{-A}}{E_{P_0}[e^{R-\Gamma}\mid E]}.
\end{equation}
The conditional formulas apply when rejection has positive probability; the corresponding unconditional integrals also cover a zero-probability rejection event. A positive mean of $R-\Gamma$ is insufficient for nominal control; even Jensen's weaker log-mean bound would require a mean of at least $\log\{1/(1-\beta)\}$ after dropping non-rejection wealth.

For a positive directional tilt, completing the square in each cohort's Gaussian density shifts the treatment mean by $\delta^*w_t$ and the control mean by $-\delta^*w_c$, where
\[
w_t=
\frac{\sigma_t^2/\Delta n_{t,k}}
{\sigma_t^2/\Delta n_{t,k}+\sigma_c^2/\Delta n_{c,k}},
\qquad w_c=1-w_t.
\]
For $k=0$, use burn-in counts. Equal maturation fractions make these weights constant. The negative direction reverses the shifts; the two-sided law chooses one sign for the entire path. The treatment-control difference shifts by $\delta^*$, but the control baseline under $Q^*$ also changes. Because the deployed target uses that baseline, $Q^*(E)$ is generally not ordinary treatment-only power.

For completeness, if $\Gamma_m=C_m-C_m^*$ and $h\ge0$, a deployed crossing at a look with $\Gamma_m\le h$ implies $M_m^*\ge e^{A-h}$. Ville's inequality therefore gives
\begin{equation}
\label{eq:slack_bound}
P_0(E)\le e^{h-A}
+P_0\!\left(\max_{1\le m\le K}\Gamma_m>h\right).
\end{equation}
This bound still depends on the unknown oracle gap. Replacing $A=\log16$ by $\log20$ controls the reference crossing probability at 5\%, but does not remove the second term for the deployed score.

The accounting identity and nominal-error criterion in Section~\ref{subsec:compensator_main} are exact under the Gaussian oracle assumptions. They do not give a finite-sample guarantee for the deployed test on non-Gaussian outcomes. The oracle baseline and variances are unknown in production, so the criterion is evaluated in simulation with known population parameters rather than as an observable validity certificate during deployment.

\subsection{Fixed-Look Local Limit and Brownian Calibration}
\label{app:local_proof}

We extend the standard fixed-look Brownian approximation to SPRT-z's estimated target and variances. The result shows that the deployed null rejection probability converges to that of the Brownian rule with the same boundaries and monitoring grid.

\paragraph{Design and assumptions.}
Fix $K\ge1$, allocation fractions $\rho_t,\rho_c>0$ summing to one, and cohort fractions $f_1,\ldots,f_K>0$ summing to one. Along an integer-compatible design sequence, let
\begin{equation}
\label{eq:local_design}
n_{g,0}=\rho_g\nu_N,\qquad
\Delta n_{g,k}=\rho_gf_kN,\qquad
\nu_N\to\infty,\qquad
\nu_N/N\to f_0\in[0,\infty).
\end{equation}
Observations are independent between arms and stationary i.i.d.\ within each arm, with common mean $\mu\ne0$ and variances $0<\sigma_g^2<\infty$. The arm distributions may otherwise differ. Set
\[
r_{MDE,N}=\frac{c}{\sqrt N},\qquad
\delta_N^*=\frac{c\mu}{\sqrt N},\qquad c>0,
\]
and define
\[
F_m=\sum_{k=1}^mf_k,\qquad
v=\frac{\sigma_t^2}{\rho_t}+\frac{\sigma_c^2}{\rho_c},\qquad
t_m=\frac{f_0+F_m}{1+f_0},\qquad
\Psi=c\mu\sqrt{\frac{1+f_0}{v}}.
\]
For standard Brownian motion $W$, the limiting scores are
\[
L_m^\pm=\pm\Psi W_{t_m}-\tfrac12\Psi^2t_m,
\qquad
L_m^{(2)}=\log\cosh(\Psi W_{t_m})-\tfrac12\Psi^2t_m.
\]
Both rules check $B<A$ at looks $1,\ldots,K$, including boundary decisions at $K$; paths with no crossing are truncated without rejection. To define the deployed rule on all paths, an invalid combined variance denominator gives score zero and no boundary decision, with non-rejection at the horizon. Consistency makes the probability of any such look tend to zero.

\begin{theorem}[Fixed-look local limit]
\label{thm:local_oracle}
Under these assumptions, for either direction and the two-sided score,
\[
\max_{1\le m\le K}|C_{m,N}-C_{m,N}^*|\xrightarrow{p}0,
\qquad
P_0(E_N)\longrightarrow p_\infty:=P_0(E_\infty),
\]
where $E_N$ and $E_\infty$ are the deployed and Brownian rejection events. If both arms have finite fourth moments, then
\[
\max_{1\le m\le K}|C_{m,N}-C_{m,N}^*|
=O_p(\nu_N^{-1/2}+N^{-1/2}).
\]
\end{theorem}

\begin{proof}
\renewcommand{\qedsymbol}{}
\textbf{Brownian reference.}
Apply the multivariate central limit theorem and continuous-mapping theorem \citep[Example~2.18 and Theorem~2.3]{vandervaart1998asymptotic} to the independent cohort sums. Since $I_{m,N}^*=(\nu_N+NF_m)/v$, their normalised cumulative differences satisfy
\[
\left(\frac{I_{m,N}^*D_{m,N}}{\sqrt{I_{K,N}^*}}\right)_{m=1}^K
\Rightarrow(W_{t_m})_{m=1}^K.
\]
This is the standard information-time representation \citep[Ch.~3]{jennison1999group}. Substituting $\delta_N^*\sqrt{I_{K,N}^*}\to\Psi$ into the oracle scores gives $(C_{m,N}^*)_{m=1}^K\Rightarrow(L_m)_{m=1}^K$.

\medskip
\noindent\textbf{SPRT-z estimation gap.}
Every lagged estimate uses at least $\rho_g\nu_N\to\infty$ observations. Consistency of sample means and variances \citep[Examples~2.1 and~2.9]{vandervaart1998asymptotic} therefore gives
\[
\frac{\psi_{m,N}}{\psi_{m,N}^*}
=\frac{\bar x_{c,m-1}}{\mu}\sqrt{\frac{I_{m,N}}{I_{m,N}^*}}
\xrightarrow{p}1,
\qquad
z_{m,N}-z_{m,N}^*
=z_{m,N}^*\left(\sqrt{\frac{J_{m,N}}{I_{m,N}^*}}-1\right)
\xrightarrow{p}0.
\]
Because $\psi_{m,N}^*,z_{m,N}^*=O_p(1)$, the directional gap satisfies, suppressing indices,
\[
\Gamma^\pm
=\pm\{(\psi-\psi^*)z^*+\psi(z-z^*)\}
-\tfrac12(\psi-\psi^*)(\psi+\psi^*)
=o_p(1).
\]
Equation~\ref{eq:two_gap} gives $|\Gamma^{(2)}|\le\max(|\Gamma^+|,|\Gamma^-|)$. With $K$ fixed, these bounds hold for the maximum gap and imply joint convergence of the deployed scores to $(L_m)_{m=1}^K$.

Under finite fourth moments, the sample-moment central limit theorem and delta method \citep[Theorem~3.1 and Example~3.2]{vandervaart1998asymptotic} give
\[
\max_m|\psi_{m,N}-\psi_{m,N}^*|
=O_p(\nu_N^{-1/2}+N^{-1/2}),\qquad
\max_m|z_{m,N}-z_{m,N}^*|=O_p(N^{-1/2}).
\]
The same gap formulas prove the stated rate.

\medskip
\noindent\textbf{Stopping decision.}
The rejection set is
\[
\mathcal E=\bigcup_{m=1}^K
\{x:B<x_j<A\text{ for all }j<m,\ x_m\ge A\}.
\]
For either score, each equation $L_m=A$ or $L_m=B$ has at most two solutions in the continuously distributed $W_{t_m}$. Hence $\mathcal E$ is a continuity set for the limiting distribution. The continuity-set form of the Portmanteau lemma \citep[see][Lemma~2.2(vii)]{vandervaart1998asymptotic} therefore gives $P_0(E_N)\to P_0(E_\infty)$, including crossings at $K$.
\end{proof}

The rate describes the score gap: for example, $\nu_N\asymp\sqrt N$ gives $O_p(N^{-1/4})$. Finite variance alone supplies convergence without a universal rate.

\paragraph{Brownian calibration.}
For the Brownian rejection event, define
\[
q_\infty=Q_\infty(E_\infty),\qquad
u_\infty=E_{Q_\infty}[e^{-R_\infty}\mid E_\infty],
\]
where $Q_\infty$ is the Brownian likelihood-ratio law and $R_\infty=L_{T_\infty}-A$ is the overshoot at rejection. Applying Appendix~\ref{app:type_i_accounting} with zero score gap gives
\[
p_\infty=\frac{\alpha}{1-\beta}q_\infty u_\infty,
\qquad
p_\infty\le\alpha\ \Longleftrightarrow\ q_\infty u_\infty\le1-\beta.
\]
Discrete crossings have strictly positive overshoot almost surely, so $0<u_\infty<1$. Thus exact tilted-law calibration to $q_\infty=1-\beta$, when attainable, yields $p_\infty=\alpha u_\infty<\alpha$.

\label{app:comp_scope}
The theorem establishes convergence of the deployed Type I error to this calibrated Brownian value along the stated design sequence. Appendix~\ref{app:validity_audits} evaluates finite-sample accuracy for the outcome distributions and monitoring schedules used in practice.

\section{Validity Audits and Stress Tests}
\label{app:validity_audits}

We evaluate the deployed rule and its reference on shared data streams, first under a stationary baseline and then under distributional and monitoring changes. These studies measure the scope of the approximations used in Appendix~\ref{app:compensator}.

\subsection{Stationary Audit and Approximation Limits}
\label{app:comp_audit}

We checked the Gaussian accounting identity on independent null and tilted-law simulations with exactly proportional arm counts, fixed true variances, 12 decision looks, and the deployed two-boundary stop. Each Gaussian cell used 20,000 paths. Direct null rejection and the independently weighted identity were statistically compatible within Monte Carlo uncertainty for both one- and two-sided tests. A reciprocal estimate based on null-path reference wealth was much less precise because a small fraction of paths carried most of the weight. These checks validate the implementation of Equation~\ref{eq:accounting} in its specified Gaussian setting; they do not make the deployed score a martingale.

For a 12-look Brownian rule with standardized terminal target 3.3, independent path simulation and deterministic grid convolution agree on ideal null rejection near 2.9\% one-sided and 2.7\% two-sided at the stated standardized target. The relevant quantity is the joint product $q_\infty u_\infty$ in Equation~\ref{eq:brownian_accounting}, rather than tilted-law rejection $q_\infty$ alone. These are numerical checks for that grid and target, not finite-sample calibration of the Negative Binomial experiments.

\begin{figure}[htbp]
    \centering
    \includegraphics[width=\linewidth]{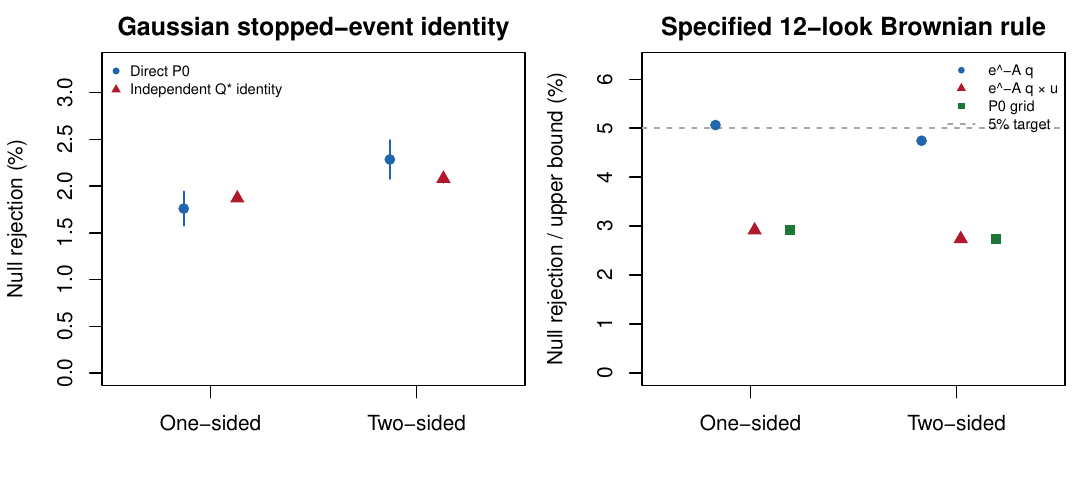}
    \caption{Gaussian and Brownian checks for the deployed stopping event and specified 12-look ideal rule. Left: direct null rejection and the independently $Q^*$-weighted identity, each with a 95\% Monte Carlo interval from 20{,}000 paths per law and side. Right: $e^{-A}q_\infty$, $e^{-A}q_\infty u_\infty$ from 100{,}000 tilted Brownian paths per side, and direct null rejection from deterministic grid convolution at spacing 0.001. The $q_\infty$-only value omits overshoot. Gaussian paths use 40 burn-in observations per arm, 50 new observations per arm per look, true arm variances 16 and 25, and fixed target 0.6. The Brownian rule uses standardized terminal target 3.3 and no burn-in information.}
    \label{fig:oracle_accounting}
\end{figure}

Figure~\ref{fig:oracle_accounting} shows the independent Gaussian estimates and the Brownian accounting. The Brownian $e^{-A}q_\infty$ values of 5.07\% and 4.75\% omit overshoot; including it gives 2.92\% and 2.74\%, agreeing with grid-convolution rejection of 2.92\% and 2.74\% at grid spacing 0.001. The one-sided $q_\infty=0.810$ exceeds $1-\beta=0.8$, while the joint product is 0.467. These values describe the specified ideal rule, not the finite-sample deployed test.

On saved stationary Negative Binomial streams, paired replay retains the realised counts and stopping events. All eighteen cell-by-side comparisons have unequal maturation fractions, so the cumulative oracle expression is an empirical comparator rather than an exact martingale. The target and scale gaps have both signs in their path tails. In one two-sided cell, mean overshoot at rejection is smaller than the mean gap; a blanket claim that overshoot exceeds the gap across the audit is therefore unsupported. A separate fixed-$K$ ladder shows declining high quantiles of the maximum score gap as traffic grows for Gaussian, Negative Binomial, and unequal-shape lognormal arms. Its finite sample rejection rates have not all approached the Brownian limit. These diagnostics illustrate the local theorem without supplying a finite-$N$ error bound.

\begin{figure}[htbp]
    \centering
    \includegraphics[width=\linewidth]{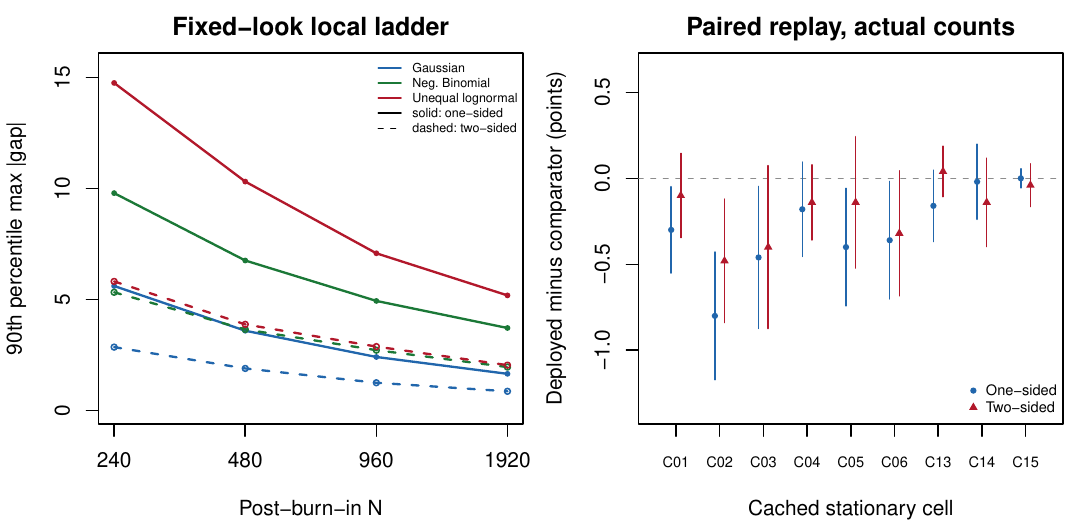}
    \caption{Fixed-look and replay diagnostics. Left: the 90th percentile of the pathwise maximum absolute deployed--oracle score gap at four post-burn-in sample sizes, with 2{,}500 independent paths per DGP, side and size; burn-in grows from 62 to 176 total observations and the relative target shrinks as $N^{-1/2}$. Right: paired deployed-minus-cumulative-comparator null rejection differences and 95\% Monte Carlo intervals on 5{,}000 saved streams per stationary Negative Binomial cell and side. Every replay stream has unequal arm maturation fractions, so this comparator has no exact martingale interpretation.}
    \label{fig:oracle_convergence_replay}
\end{figure}

Figure~\ref{fig:oracle_convergence_replay} displays the gap trend and paired replay differences. At $N=1{,}920$, the unequal-shape lognormal two-sided deployed rejection is 4.76\% against 2.72\% for its comparator, a paired difference of 2.04 percentage points (standard error 0.47). The one-sided comparison goes in the opposite direction. The finite-$N$ curves do not show uniform convergence of rejection rates.

\subsection{Common-Arm and Differential-Arm Stress Tests}
\label{app:comp_attack}

Table~\ref{tab:adversarial} compares methods on identical streams in two-sided tests. The common-arm battery keeps the arms identically distributed; the differential-arm battery holds cohort means equal while changing arm shapes.

\textbf{Common-arm generators.} The baseline follows Section~\ref{sec:simframework}, with $n_{daily}=500$ and $r_{MDE}=0.10$. Static contamination multiplies 1\% of units' outcomes by 25; burst contamination applies the same multiplier to 2\% of units enrolling after one third of the simulated enrolment period. At the same onset, the changepoint raises both arms' daily outcomes by 50\%, and the allocation attack changes assignment from 50/50 to 80/20. The drift cell compounds both arms' daily means by 4\%. Near-continuous monitoring uses $n_{daily}=100$, $r_{MDE}=0.20$, and about 236 available looks. The extreme-skew cell uses log standard deviation 1.4, Negative Binomial dispersion 0.5, and $r_{MDE}=0.20$. Contamination, drift, changepoint, and allocation attacks retain the clean-baseline calibration; extreme skew is calibrated on its own stationary generator.

\textbf{Differential-arm generators.} Tests use 500 units per look, fixed arm counts, $r_{MDE}=0.10$, and a 17,500-unit horizon. The stationary Negative Binomial generator draws a latent rate $L$ with $\ln L\sim N(\ln3,1.1^2)$ and a count with mean $7L$ and size 7. The other generators have mean 40 in each arm. Normal arms have standard deviations 20 and 60; Student-$t_3$ arms use scales 12 and 36. Lognormal arms have log standard deviations 0.7 and 1.4; Gamma arms have shapes 4 and 0.5. The high- and low-dispersion generators use a lognormal control $B$ with log standard deviation 0.8 and treatment $40B^q/\mathbb E[B^q]$, with $q=1.8$ and $0.45$, respectively. The time-varying generator alternates these treatment shapes by cohort. Allocation is 50/50 except in the 80/20 Normal cell.

\textbf{Comparators.} GST uses Pampallona--Tsiatis boundaries with O'Brien--Fleming-type $\alpha$ and $\beta$ spending, binding futility, and $\Delta=0$ \citep{pampallona1994}. In the common-arm battery it uses its own horizon and at most twenty analyses. In the differential-arm battery it uses 17 analyses over the shared horizon; SPRT-z and mSPRT have 34 evaluable looks after initialisation. FHT uses the size planned from control variance, 1,000--9,000 units. An own-sized GST is omitted there because some generators plan fewer than two looks. The efficacy-only mSPRT uses a normal-mixture scale at the absolute MDE and the same truncation horizon as SPRT-z \citep{johari2022}.

\begin{table}[ht]
    \caption{Stress tests: realised Type I error (\%) of the as-shipped two-boundary truncated configuration, by scenario and statistic, every method on identical streams. Common-arm rows: 5{,}000 replications (2{,}000 near-continuous), group sequential design at its own horizon, fixed-horizon test at $N_{FHT}$. Differential-arm rows: 5{,}000 replications at the shipped inputs $(a,b) = (0.05, 0.20)$, group sequential design with 17 analyses on the shared 17{,}500-unit horizon, fixed-horizon test at the size planned from the control arm's variance; the Recalibrated column reruns the deployed form at $(a,b) = (0.02, 0.05)$ on 10{,}000 fresh streams. Bold marks rates above the 5\% target beyond the marginal Monte Carlo half-width (0.5 to 1.1 points in the common-arm rows, 0.6 in the differential-arm rows).}
    \label{tab:adversarial}
    \centering
    \footnotesize
    \setlength{\tabcolsep}{3pt}
    \begin{tabular}{lrrrrr}
        \toprule
        \textbf{Scenario} & \textbf{Deployed} & \textbf{GST} & \textbf{mSPRT} & \textbf{FHT} & \textbf{Recalibrated} \\
        \midrule
        \multicolumn{6}{l}{\textit{Common-arm attacks (arms identically distributed)}} \\
        Stationary control        & 3.3 & \textbf{5.7} & 1.4 & 5.3 & \\
        Whales, static            & 0.6 & 4.2 & 0.2 & 4.1 & \\
        Whales, burst             & 3.1 & 4.8 & 1.1 & 4.9 & \\
        Changepoint $+50\%$       & 3.3 & 4.9 & 1.3 & 5.1 & \\
        Drift 4\%/day             & 3.5 & 5.1 & 1.2 & 5.3 & \\
        Allocation shift          & 3.4 & 5.3 & 1.2 & 4.9 & \\
        Near-continuous ($K \approx 236$) & 3.1 & 4.2 & 2.1 & 4.2 & \\
        Extreme skew              & 2.4 & 4.9 & 0.8 & 4.5 & \\
        \midrule
        \multicolumn{6}{l}{\textit{Differential-arm attacks (equal means, unequal shapes)}} \\
        Stationary Negative Binomial       & 2.8 & 4.9 & 1.3 & 4.7 & 1.1 \\
        Normal, variance ratio 9           & 3.3 & 4.9 & 1.9 & 5.3 & 1.4 \\
        Lognormal, unequal shape           & \textbf{7.6} & \textbf{5.9} & 4.6 & \textbf{6.7} & 3.7 \\
        Gamma, unequal shape               & 3.3 & 5.1 & 2.1 & 5.4 & 1.3 \\
        High-dispersion treatment marginal & \textbf{7.2} & \textbf{6.0} & 4.3 & \textbf{6.1} & 3.3 \\
        Low-dispersion treatment marginal  & 2.5 & 5.4 & 1.8 & 5.4 & 1.0 \\
        Time-varying heterogeneity         & \textbf{6.1} & \textbf{6.1} & 4.1 & \textbf{6.1} & 2.9 \\
        80/20 allocation, variance ratio 9 & 2.9 & 5.1 & 1.7 & 5.1 & 1.2 \\
        Student-$t_3$, variance ratio 9    & 2.7 & 4.7 & 1.5 & 5.4 & 1.0 \\
        \bottomrule
    \end{tabular}
\end{table}

\textbf{Common-arm results.} Deployed rejection never exceeds 3.5\%; simultaneous Clopper--Pearson bounds put all eight true rates below 4.3\% with at least 95\% confidence. FHT and GST are near nominal, apart from GST's stationary-control rate of 5.7\%; mSPRT is conservative. Static contamination makes the deployed form particularly conservative, but reduces its power to 12.8\%, with all methods between 7.4\% and 33.2\%. Variance inflation slows information accrual beyond the clean-baseline horizon. Error control in this case accompanies a substantial detection cost.

\textbf{Unequal-shape results.} Unequal-shape Lognormal, high-dispersion treatment, and time-varying generators give deployed rejection rates of 7.60\%, 7.16\%, and 6.06\%, each above nominal beyond Monte Carlo error. GST and planned FHT also exceed nominal on these three generators. Unequal variance alone does not reproduce the excess in the Normal and Student-$t_3$ cells. Treatment-arm right skew, estimated scales and monitoring cadence are plausible contributors; these comparisons do not isolate their effects. Exact Gaussian accounting does not cover these finite-sample settings.

On GST's 17-look schedule, SPRT-z gives 5.72\%, 5.92\%, and 4.90\%, showing that cadence contributes materially to the excess. mSPRT stays below nominal throughout; FHT at the full 17,500-unit horizon is near nominal (4.52\% to 5.58\%). These comparisons use the same 5,000 streams per generator; the marginal 95\% Monte Carlo half-width near 5\% rejection is 0.60 percentage points.

\textbf{Recalibration and power.} Tuning over $a\in\{0.015,0.018,0.020,0.022,0.025,0.030,0.050\}$, $b\in\{0.05,0.10,0.15,0.20\}$, and horizons from 20 to 45 looks uses 5,000 replications per generator. It selects boundary inputs $(a,b)=(0.02,0.05)$, corresponding to $(A,B)=(3.861,-2.975)$, at 35 looks including initialisation and 17,500 units. Validation on 10,000 fresh replications per generator gives a maximum rejection rate of 3.70\% and a simultaneous 97.5\% upper bound of 4.25\% across all nine generators.

At a $+10\%$ lift, fresh-seed power is 80.82\% on the stationary Negative Binomial generator, with a one-sided 97.5\% lower bound of 80.03\%. Table~\ref{tab:recal_power} extends the power comparison to all nine generators and both effect directions, using 5,000 replications per cell on the tuning seeds. Power reaches 85\% to 97\% on six other generators, but only 68\% and 59\% on the two heavy-treatment-skew generators. FHT at the same horizon and $\alpha$ input reaches 88\% and 82\% on those two. Thus restored Type I control does not ensure adequate power. Power differs between the effect directions; multiplying treatment outcomes by $1\pm0.10$ changes both their mean and variance.

These inputs are calibrated for the tested metric family, target, cadence, and horizon. Deployments exposed to treatment-arm skew should validate against differential-arm nulls and reassess the horizon when treatment variance rises.

\begin{table}[H]
\caption{Power (\%) at the recalibrated inputs $(a,b) = (0.02, 0.05)$, 35 looks, 17{,}500-unit horizon, 5{,}000 replications per cell. The fixed-horizon test reads the full horizon. The two heavy-treatment-skew generators are the two on which the deployed form falls below the 80\% target; the fixed-horizon test at the same size and $\alpha$ input has higher power on both, showing that recalibrating Type I error does not ensure adequate sequential power.}
\label{tab:recal_power}
\centering
\footnotesize
\begin{tabular}{lrrrrrr}
\toprule
& \multicolumn{3}{c}{Lift $+10\%$} & \multicolumn{3}{c}{Lift $-10\%$} \\
\cmidrule(lr){2-4} \cmidrule(lr){5-7}
\textbf{Generator} & Deployed & GST & FHT & Deployed & GST & FHT \\
\midrule
Stationary Negative Binomial       & 80.6 & 89.2 & 91.7 & 88.5 & 95.1 & 96.6 \\
Normal, variance ratio 9           & 95.3 & 99.7 & 99.8 & 97.0 & 100.0 & 100.0 \\
Lognormal, unequal shape           & 68.1 & 83.3 & 87.6 & 87.8 & 90.4 & 91.9 \\
Gamma, unequal shape               & 95.4 & 99.9 & 99.9 & 97.4 & 100.0 & 100.0 \\
High-dispersion treatment marginal & 59.1 & 77.3 & 81.9 & 82.8 & 86.2 & 87.3 \\
Low-dispersion treatment marginal  & 96.5 & 100.0 & 100.0 & 97.6 & 100.0 & 100.0 \\
Time-varying heterogeneity         & 85.3 & 97.2 & 98.6 & 94.2 & 96.7 & 96.5 \\
80/20 allocation, variance ratio 9 & 96.2 & 100.0 & 100.0 & 97.7 & 100.0 & 100.0 \\
Student-$t_3$, variance ratio 9    & 95.8 & 99.6 & 99.7 & 96.9 & 99.9 & 99.9 \\
\bottomrule
\end{tabular}
\end{table}

\section{Efficiency Comparisons with GST and mSPRT}
\label{app:comparisons}

\subsection{GST under Matched Error Targets and a Shared Horizon}
\label{app:comp_matched}

Nominal boundary inputs give SPRT-z and GST different realised error rates. We calibrate each method's $\alpha$ input on the stationary DGP at $n_{daily}=500$, using disjoint calibration and evaluation streams and 5,000 fresh shared replications per effect. The inputs targeting 5\% Type I error are 0.077 for SPRT-z and 0.048 for Pampallona--Tsiatis GST. Table~\ref{tab:matched_comparison} distinguishes two comparisons.

\begin{table}[H]
\centering
\small
\caption{GST comparisons under matched error targets and a shared horizon. Each method targets 5\% Type I error. The matched-error design also targets 80\% MDE power using its own horizon. Rates and expected sample sizes are measured on fresh streams; all sample sizes are totals across arms.}
\label{tab:matched_comparison}
\begin{tabular}{lrr}
\toprule
 & SPRT-z & GST \\
\midrule
\multicolumn{3}{l}{\textit{Matched error targets, own horizons}} \\
Maximum sample & 16,644 & 11,400 \\
Realised Type I error & 5.0\% & 4.5\% \\
Power at MDE & 80.3\% & 80.7\% \\
Expected sample under null & 5,826 & 6,613 \\
Expected sample at MDE & 6,400 & 7,304 \\
\midrule
\multicolumn{3}{l}{\textit{Shared horizon, Type I error calibrated}} \\
Maximum sample & 15,660 & 15,660 \\
Power at MDE & 79.8\% & 90.6\% \\
Expected sample under null & 5,810 & 8,957 \\
\bottomrule
\end{tabular}
\end{table}

With matched error targets, SPRT-z uses about 12\% less expected sample at both design hypotheses and 4\% to 10\% less at intermediate effects. It uses less throughout the evaluated grid from $-$MDE to $1.5\times$MDE. FHT uses 9,140 units. Under spike-and-slab weights with null mass 0.5 to 0.9 and the remainder at the MDE or split with MDE$/2$, SPRT-z uses 64\% to 68\% of FHT's expected sample, versus 73\% to 76\% for GST; ship rates differ by at most 0.6 points. At the shared horizon, GST instead achieves higher power and uses more expected sample under the null. These findings cover one DGP, cadence, and boundary family.

\subsection{mSPRT, Futility, and Effects above the MDE}
\label{app:msprt_comparison}

In the common-arm stress comparison, efficacy-only mSPRT uses about 2.5 times SPRT-z's expected null sample. Adding a confidence-sequence futility rule that excludes the MDE reduces the ratio to 1.8 at the most generous evaluated budget, $\alpha_{fut}=\beta$.

Table~\ref{tab:msprt_grid} extends the comparison across effect sizes and cadences on the Negative Binomial DGP of Section~\ref{sec:simframework}, using 5,000 replications per cell. Each method uses its own power-calibrated horizon, inputs $(\alpha,\beta)=(0.05,0.20)$, and a relative design MDE of 0.10; the mSPRT normal-mixture scale equals the absolute MDE. At $n_{daily} = 500$, SPRT-z's expected sample is 0.71 of efficacy-only mSPRT's at the MDE and 0.92 at twice the MDE. It crosses parity near three times the MDE (1.03) and reaches 1.11 at five times, where power is saturated at 100\% for both; the fine cadence behaves similarly. The mixture distributes evidence across effect sizes, whereas the point-alternative likelihood ratio accumulates at a rate set by the MDE.

At $n_{daily} = 2{,}500$, the ratio reaches 2 at five times the MDE. SPRT-z uses its first look to initialise the prequential baseline, so it cannot stop before the second look at about 5{,}000 units and its expected sample floors near 6{,}400. This cost comes from relative-MDE initialisation at coarse cadence. At $n_{daily}=500$ and half the MDE, futility stopping reduces SPRT-z's expected sample to about 7{,}600 units, versus 12{,}800 for futility-equipped mSPRT. At the finer cadences, its advantage lies between zero and about twice the business threshold.

\begin{table}[ht]
\centering
\small
\caption{Expected sample size of SPRT-z divided by that of efficacy-only mSPRT, with each method using its own power-calibrated horizon. Ratios below one favour SPRT-z. All tests are two-sided, with 5,000 replications per cell; power is 100\% for both methods at effects of twice the MDE and above.}
\label{tab:msprt_grid}
\begin{tabular}{lrrr}
\toprule
True effect ($\times$MDE) & $n_{daily}=100$ & $n_{daily}=500$ & $n_{daily}=2500$ \\
\midrule
0   & 0.36 & 0.39 & 0.50 \\
0.5 & 0.48 & 0.52 & 0.66 \\
1   & 0.68 & 0.71 & 0.82 \\
1.5 & 0.84 & 0.84 & 1.00 \\
2   & 0.93 & 0.92 & 1.25 \\
3   & 1.05 & 1.03 & 1.76 \\
5   & 1.17 & 1.11 & 1.99 \\
\bottomrule
\end{tabular}
\end{table}

\section{Scale-Free Horizon Calibration (SFHC) Mechanics}
\label{app:sfhc}

Continuous variance updates make the power-preserving horizon $N_{max}$ analytically intractable for a discrete, prequential SPRT. SFHC turns the standard Brownian mapping of sequential statistics \citep{jennison1999group, grunwald2024} into a computational calibration.

It uses Monte Carlo bisection on the standardized $Z$-scale to choose $N_{max}$ for Type II error near $\beta$ under a specified monitoring schedule, independently of metric variance.

\subsection{Information Time and Scale-Free Drift}
Let $N_{FHT}$ be the fixed-horizon requirement, $N_{max}$ the candidate sequential horizon, and $n_{daily}$ the batch size. The calibration uses $K=\lceil N_{max}/n_{daily}\rceil$ looks at information fractions $t_k=k/K$ and $S$ simulated paths. The matured-cohort process is approximated by Brownian motion on this information scale.

For a powered two-sided FHT, the design score and candidate-horizon drift are
\[
Z_{FHT}=\Phi^{-1}(1-\alpha/2)+\Phi^{-1}(1-\beta),
\qquad
\Psi_{max}=Z_{FHT}\sqrt{N_{max}/N_{FHT}}.
\]
Here $\Phi^{-1}$ is the standard normal quantile. The second relation follows from $Z_{FHT}=c\sqrt{N_{FHT}}$ and $\Psi_{max}=c\sqrt{N_{max}}$: taking their ratio cancels the metric-specific signal-to-noise ratio $c=\delta/\sigma$. Calibration inputs and realised error rates remain distinct, as in Section~\ref{subsec:boundaries}.

\subsection{Algorithm Mechanics: Simulation and Bisection Search}
Algorithm~\ref{alg:sfhc} bisects over a Brownian simulation to find $N_{max}$ with power $1-\beta$.

\begin{algorithm}[H]
    \caption{SFHC Bisection Search for Maximum Horizon ($N_{max}$)}
    \label{alg:sfhc}
    \centering
    \small
    \begin{tabular}{ll}
        \toprule
        \multicolumn{2}{l}{\textbf{Require:} $N_{FHT}$, $n_{daily}$, $\alpha$, $\beta$, tolerance $\epsilon$, max iterations $J$, simulations $S$} \\
        \multicolumn{2}{l}{\textbf{Ensure:} Calibrated horizon $N_{max}$ yielding power $\approx 1 - \beta$} \\
        \midrule
        1: & $L \leftarrow N_{FHT}$ \\
        2: & $U \leftarrow 2 \cdot N_{FHT}$ \\
        3: & $Z_{FHT} \leftarrow \Phi^{-1}(1 - \alpha/2) + \Phi^{-1}(1 - \beta)$ \\
        4: & Calculate Wald boundaries: $A = \ln(\frac{1 - \beta}{\alpha})$, $B = \ln(\frac{\beta}{1 - \alpha})$ \\
        5: & \textbf{for} $j=1, \dots, J$ \textbf{do} \\
        6: & \quad $N_{mid} \leftarrow (L + U) / 2$ \\
        7: & \quad $K \leftarrow \lceil N_{mid} / n_{daily} \rceil$ \\
        8: & \quad $\Psi_{max} \leftarrow Z_{FHT} \sqrt{N_{mid} / N_{FHT}}$ \\
        9: & \quad $hits \leftarrow 0$ \\
        10: & \quad \textbf{for} $s=1, \dots, S$ \textbf{do} \quad \textit{$\triangleright$ Evaluate statistical power} \\
        11: & \quad \quad Generate standard Brownian path $B(t_k)$ for $k=1 \dots K$, with $t_k = k/K$ \\
        12: & \quad \quad $Z_{t_k} \leftarrow \frac{B(t_k) + \Psi_{max} \cdot t_k}{\sqrt{t_k}}$ \\
        13: & \quad \quad Calculate $C_k^{(2)} = \ln \left( \frac{\phi(Z_{t_k} - \Psi_{max} \sqrt{t_k}) + \phi(Z_{t_k} + \Psi_{max} \sqrt{t_k})}{2 \phi(Z_{t_k})} \right)$ \\
        14: & \quad \quad \textbf{if} path crosses $A$ prior to crossing $B$ \textbf{then} \\
        15: & \quad \quad \quad $hits \leftarrow hits + 1$ \\
        16: & \quad \quad \textbf{end if} \\
        17: & \quad \textbf{end for} \\
        18: & \quad $\hat{P} \leftarrow hits / S$ \\
        19: & \quad \textbf{if} $|\hat{P} - (1 - \beta)| \le \epsilon$ \textbf{then} \textbf{break} \\
        20: & \quad \textbf{if} $\hat{P} < 1 - \beta$ \textbf{then} $L \leftarrow N_{mid}$ \textbf{else} $U \leftarrow N_{mid}$ \\
        21: & \textbf{end for} \\
        22: & \textbf{return} $N_{mid}$ \\
        \bottomrule
    \end{tabular}
\end{algorithm}

For candidate $N_{max}$, set $K = \lceil N_{max} / n_{daily} \rceil$ and $t_k = k / K$. Draw independent increments $\Delta B_k \sim \mathcal{N}(0, \Delta t)$ with $\Delta t = 1 / K$, and accumulate $B(t_k) = \sum_{i=1}^k \Delta B_i$. All candidates share a standard-normal matrix spanning the largest look count in the search; each uses its first $K$ rows scaled to $\Delta t = 1/K$. This gives the correct marginal Brownian law but does not nest candidates on one Brownian path. Near-monotonicity is therefore measured, not guaranteed by construction.

Diagnostics use sixty candidate horizons and $10{,}000$ simulations per candidate on the main-study anchor at all three cadences. Fresh-noise curves have $12$ to $19$ non-monotone decrements of up to $1.8$ points; row rescaling has at most $4$ of up to $0.5$ points. Nested-path coupling has at most $5$ of up to $0.9$ points. Both coupled versions select the same canonical horizon. Across twenty independent calibrations, selected $N_{max}$ has standard deviation $250$ to $570$ units, about $2$ to $4$ percent. Validation at each selected horizon with $200{,}000$ fresh paths gives power within $0.3$ points of $80\%$ in every tested configuration. Production calibration should include this independent validation.

\section{Brownian Median Unbiased Estimator (MUE) Mechanics}
\label{app:mue}

The MUE and CIs invert boundary-crossing probabilities under stagewise ordering.

We describe the two-sided implementation used in the simulations. The observed outcome is the stopping look $m^*$, final statistic $z^*$, and decision $D\in\{H_1,H_0,T\}$, where $T$ denotes truncation without a boundary crossing. Write $\Omega$ for the outcome space, $M$ for the maximum number of simulation looks, $\Psi$ for a candidate true drift, and $S$ for the number of Brownian paths. Information fractions are denoted by $t_m$, distinct from the inverse-variance information $I_m$ in Appendix~\ref{app:compensator}.

\subsection{Sample Space Partition and Stagewise Ordering}
The terminal outcome is $(m^*,z^*,D)$. We partition $\Omega$ into six regions by decision and the sign of $z^*$, ordered from most positive to most negative:
\begin{align*}
    \mathcal{R}_6 &= \{ (m^*, z^*, D) : D = H_1 \text{ and } z^* > 0 \} \\
    \mathcal{R}_5 &= \{ (m^*, z^*, D) : D = T \text{ and } z^* > 0 \} \\
    \mathcal{R}_4 &= \{ (m^*, z^*, D) : D = H_0 \text{ and } z^* > 0 \} \\
    \mathcal{R}_3 &= \{ (m^*, z^*, D) : D = H_0 \text{ and } z^* \le 0 \} \\
    \mathcal{R}_2 &= \{ (m^*, z^*, D) : D = T \text{ and } z^* \le 0 \} \\
    \mathcal{R}_1 &= \{ (m^*, z^*, D) : D = H_1 \text{ and } z^* \le 0 \}
\end{align*}

Write $o\succeq o^*$ when simulated outcome $o=(m,z,D)$ ranks at least as positive as the observed outcome $o^*=(m^*,z^*,D^*)$. An outcome in a higher-numbered region always ranks above one in a lower-numbered region. Within the same region, use the following comparisons:
\begin{center}
\begin{tabular}{ll}
\toprule
Region of both outcomes & Condition for $o\succeq o^*$ \\
\midrule
$\mathcal R_6$ (positive efficacy), $\mathcal R_3$ (negative futility)
 & $m<m^*$ or $(m=m^*\text{ and }z\ge z^*)$ \\
$\mathcal R_4$ (positive futility), $\mathcal R_1$ (negative efficacy)
 & $m>m^*$ or $(m=m^*\text{ and }z\ge z^*)$ \\
$\mathcal R_5$, $\mathcal R_2$ (truncation)
 & $z\ge z^*$ \\
\bottomrule
\end{tabular}
\end{center}
Thus stopping-time comparisons reverse between positive and negative outcomes, while larger signed $z$ breaks every same-look tie. Truncated paths all stop at $M$. Equality is included in the tail probability.

\subsection{Brownian Simulation and Common Random Numbers (CRN)}
Set $M=\lceil N_{max}/n_{daily}\rceil$ and $t_m=m/M$. Draw independent increments $\Delta B_m^{(j)}\sim N(0,1/M)$ for $m=1,\ldots,M$ and paths $j=1,\ldots,S$, and accumulate $B^{(j)}(t_m)$. The fixed design drift and targets are
\[
\Psi_{design}=\left[\Phi^{-1}(1-\alpha/2)+\Phi^{-1}(1-\beta)\right]
\sqrt{N_{max}/N_{FHT}},\qquad
\psi_{design,m}=\Psi_{design}\sqrt{t_m}.
\]
For a candidate true drift $\Psi_{guess}$, simulate
\[
Z_m^{(j)}=\frac{B^{(j)}(t_m)+\Psi_{guess}t_m}{\sqrt{t_m}},\qquad
L_m^{(j)}=\ln\cosh\!\left(\psi_{design,m}Z_m^{(j)}\right)
-\tfrac12\psi_{design,m}^2.
\]
Only the simulated data drift changes during inversion; the design targets, boundaries $A=\ln((1-\beta)/\alpha)$ and $B=\ln(\beta/(1-\alpha))$, and horizon remain fixed.

The Brownian implementation checks both boundaries from $m=1$. It treats the design target as known and does not simulate the deployed rule's initial estimation of the prequential target. The observed $m^*$ is the daily index from the first matured-data date, including the initialisation day; it is not renumbered from the first evaluable likelihood ratio. Equally spaced information fractions also approximate the empirical, potentially uneven accrual of matured units. These conventions are part of the approximation evaluated in Section~\ref{sec:bias}.

Stop each simulated path at its first boundary crossing, or at $M$ if none occurs. A crossing at $M$ is classified as efficacy or futility; only a path with no crossing is truncated. The empirical one-sided tail probability is the proportion of simulated stopping outcomes $o_j$ satisfying $o_j\succeq o^*$.

Bisection uses an increasing tail probability in the candidate drift $\Psi_{guess}$. We reuse one Brownian increment matrix across candidates within each inversion, so changes in the estimated probability reflect the candidate effect rather than newly drawn noise. Monotonicity held on the evaluated grids, but CRN does not establish it for the six-region ordering; a general proof remains open. Separate experiments and simulation replications use independent matrices to avoid imposing a common simulation error on all estimates and intervals.

\begin{algorithm}[h]
    \caption{Unconditional P-value Computation}
    \label{alg:pvalue}
    \centering
    \small
    \begin{tabular}{ll}
        \toprule
        \multicolumn{2}{l}{\textbf{Require:} Observed outcome $o^*=(m^*,z^*,D)$; candidate drift $\Psi_{guess}$;} \\
        \multicolumn{2}{l}{\phantom{\textbf{Require:}} paths $B^{(j)}(t_m)$; fixed $\psi_{design,m}$, $A$, $B$, $M$, and $S$} \\
        \multicolumn{2}{l}{\textbf{Ensure:} Estimated p-value $\hat{p}$} \\
        \midrule
        1: & \textbf{for} each simulated path $j \in \{1, \dots, S\}$ \textbf{do} \\
        2: & \quad $Z_m^{(j)} \leftarrow [B^{(j)}(t_m)+\Psi_{guess}t_m]/\sqrt{t_m}$, $m=1,\dots,M$ \\
        3: & \quad $L_m^{(j)} \leftarrow \ln\cosh(\psi_{design,m}Z_m^{(j)})-\psi_{design,m}^2/2$ \\
        4: & \quad $u_j\leftarrow\min\{m:L_m^{(j)}\ge A\}$; $\ell_j\leftarrow\min\{m:L_m^{(j)}\le B\}$ \\
        5: & \quad Set a missing crossing to $\infty$; stop at $s_j\leftarrow\min(u_j,\ell_j,M)$ \\
        6: & \quad Set $D_j=H_1$ if $u_j\le\min(\ell_j,M)$, $H_0$ if $\ell_j< u_j$ and $\ell_j\le M$; else $T$ \\
        7: & \quad Form $o_j=(s_j,Z_{s_j}^{(j)},D_j)$ and compare $o_j\succeq o^*$ \\
        8: & \textbf{end for} \\
        9: & \textbf{return} $\hat{p} = S^{-1}\sum_{j=1}^{S}\mathbbm{1}\{o_j\succeq o^*\}$ \\
        \bottomrule
    \end{tabular}
\end{algorithm}

\subsection{Bisection Search for Estimates}
Algorithm~\ref{alg:bisection} inverts the simulated tail probability over $\Psi_{guess}$. The implementation uses $[\Psi_L,\Psi_U]=[-10,15]$ and 25 iterations in the fast routine (40 in the direct routine), reusing the same Brownian paths for the MUE and both endpoints. It does not expand the bracket or signal an unbracketed target: if $p^*$ lies outside the endpoint probabilities, the iteration approaches a search limit. Such an output is not a resolved inversion; endpoint probabilities should be checked, and an unbracketed target requires a wider search. These numerical limits are not confidence limits on the effect.

\begin{algorithm}[h]
    \caption{Bisection Search for MUE and CI Bounds}
    \label{alg:bisection}
    \centering
    \small
    \begin{tabular}{ll}
        \toprule
        \multicolumn{2}{l}{\textbf{Require:} Target $p^*$; bounds $\Psi_L,\Psi_U$; iterations $J$; fixed inputs to Algorithm~\ref{alg:pvalue}} \\
        \multicolumn{2}{l}{\textbf{Precondition:} $\hat p(\Psi_L)\le p^*\le\hat p(\Psi_U)$ and an increasing tail probability} \\
        \multicolumn{2}{l}{\textbf{Ensure:} Numerical drift estimate $\hat{\Psi}$ for the target probability} \\
        \midrule
        1: & \textbf{for} $j=1, \dots, J$ \textbf{do} \\
        2: & \quad $\Psi_{mid} \leftarrow (\Psi_L + \Psi_U) / 2$ \\
        3: & \quad $\hat{p} \leftarrow$ \textsc{ComputePValue}($\Psi_{mid}$) \quad \textit{$\triangleright$ via Algorithm~\ref{alg:pvalue}} \\
        4: & \quad \textbf{if} $\hat{p} > p^*$ \textbf{then} \\
        5: & \quad \quad $\Psi_U \leftarrow \Psi_{mid}$ \\
        6: & \quad \textbf{else} \\
        7: & \quad \quad $\Psi_L \leftarrow \Psi_{mid}$ \\
        8: & \quad \textbf{end if} \\
        9: & \textbf{end for} \\
        10: & \textbf{return} $(\Psi_L + \Psi_U) / 2$ \\
        \bottomrule
    \end{tabular}
\end{algorithm}

Use $p^*=0.025$, $0.500$, and $0.975$ for the lower 95\% endpoint, MUE, and upper endpoint, respectively.

\subsection{Algebraic Conversion to the Absolute Metric}
Bisection returns standardized parameters, which we map to absolute $\delta$ using $t_{stop}=n_{stop}/N_{max}$, where $n_{stop}$ is the total observed sample across arms. This conversion uses the realised sample fraction; the Brownian ordering uses the daily index $m^*$ on the grid $t_m=m/M$.

Equating expected Brownian and empirical $Z$ scores gives
$$\delta_{MUE} = \Psi_{MUE} \cdot SE_{stop} \cdot \sqrt{t_{stop}}$$

Apply this formula to $\Psi_{upper}$, $\Psi_{MUE}$, and $\Psi_{lower}$, using the current unpooled $SE_{stop}$ from Equation~\ref{eq:zscore}. The reported relative effects divide each absolute estimate or endpoint by the observed control mean at stopping; the evaluated metrics have positive control means.

\subsection{Approximation of Prequential Drift}
The Brownian search assumes constant drift ($\Psi_{guess}$), whereas SPRT-z updates its target and variance estimates. Fixed-drift simulation is an approximation for bias correction and interval construction. Its accuracy depends on the realised burn-in, scale estimates and monitoring schedule; the local result of Appendix~\ref{app:local_proof} does not establish finite-sample MUE or interval coverage.

\section{Simulation Details for the Bernoulli DGP}\label{app:simulation_details}

We repeat the validation with a bounded DGP to separate the results from Negative Binomial tails and to study settings where $N_{FHT}$ is small relative to $n_{daily}$.

\subsection{Generative Process and Configuration}
Each unit $i$ in arm $a \in \{C, T\}$ contributes a single outcome
$$Y_{i,a} \sim \mathrm{Bernoulli}(p_a), \qquad p_C = 0.30, \qquad p_T = p_C (1 + \delta),$$
with $\delta\in\{0,0.05,0.10,0.15\}$ and equally split $n_{daily}\in\{100,500,2500\}$. Staggered cohorts match Section~\ref{sec:simframework}; each fixed window records the unit's single Bernoulli outcome on its enrolment day.

For the two-sided test, $\alpha = 0.05$ and $\beta = 0.20$ give Wald boundaries $A = \log((1-\beta)/\alpha) = 2.77$ and $B = \log(\beta/(1-\alpha)) = -1.56$. Null-simulation means and variances yield total $N_{FHT}$ values of 7{,}524, 7{,}420, and 7{,}351, within 2.3\% of the analytic two-proportion $Z$-test requirement of 7{,}520 at $p_C = 0.30$ and $\delta = 0.10$. Cadence-specific SFHC calibration (Appendix~\ref{app:sfhc}) inflates these by $1.6$ to $1.8$ times, as in the Negative Binomial study at comparable look counts.

\subsection{Results}
Table~\ref{tab:bernoulli_summary} reports 2,500 stationary simulations per $n_{daily}\times\delta$ cell. $N_{FHT}/n_{daily}$ is the number of full FHT monitoring days; smaller values mean coarser information time.

\begin{table}[H]
    \caption{Comparative Performance of FHT vs. SPRT-z under a Bernoulli DGP (2,500 Simulations, Stationary Baseline). Rejection Rate reports the empirical Type I error at $\delta = 0.00$ and the statistical power at $\delta > 0$; rates are shown with 95\% Monte Carlo half-widths. SS Reduction is $1 - \bar N_{\text{SPRT-z}} / \bar N_{\text{FHT}}$. CI Coverage is the unconditional coverage of the 95\% confidence interval. MUE Bias Reduction is the relative reduction in median bias from the raw SPRT-z estimate to the MUE-adjusted estimate.}
    \label{tab:bernoulli_summary}
    \centering
    \footnotesize
    \setlength{\tabcolsep}{3pt}
    \begin{tabular}{lrrrrrrr}
        \toprule
        & & & \multicolumn{2}{c}{\textbf{Rejection Rate}} & & & \\
        \cmidrule(lr){4-5}
        \textbf{Daily Batch} & \textbf{$\delta$} & \textbf{$N_{FHT}/n_{daily}$} & \textbf{FHT} & \textbf{SPRT-z} & \textbf{SS Reduction} & \textbf{CI Coverage} & \textbf{MUE Bias Red.} \\
        \midrule
        \textbf{$n_{daily} = 100$}
        & 0.00 (Null) & 76 & 0.052 {\scriptsize$\pm$.009} & 0.041 {\scriptsize$\pm$.008} & 39.4\% & 0.954 {\scriptsize$\pm$.008} & --- \\
        & 0.05        & 76 & 0.303 {\scriptsize$\pm$.018} & 0.264 {\scriptsize$\pm$.017} & 25.9\% & 0.963 {\scriptsize$\pm$.007} & 83.9\% \\
        & 0.10 (MDE)  & 76 & 0.812 {\scriptsize$\pm$.015} & 0.781 {\scriptsize$\pm$.016} & 32.4\% & 0.965 {\scriptsize$\pm$.007} & 98.4\% \\
        & 0.15        & 76 & 0.989 {\scriptsize$\pm$.004} & 0.972 {\scriptsize$\pm$.006} & 54.3\% & 0.943 {\scriptsize$\pm$.009} & 93.6\% \\
        \midrule
        \textbf{$n_{daily} = 500$}
        & 0.00 (Null) & 15 & 0.057 {\scriptsize$\pm$.009} & 0.033 {\scriptsize$\pm$.007} & 33.5\% & 0.955 {\scriptsize$\pm$.008} & --- \\
        & 0.05        & 15 & 0.301 {\scriptsize$\pm$.018} & 0.257 {\scriptsize$\pm$.017} & 17.8\% & 0.967 {\scriptsize$\pm$.007} & 96.3\% \\
        & 0.10 (MDE)  & 15 & 0.808 {\scriptsize$\pm$.015} & 0.793 {\scriptsize$\pm$.016} & 25.7\% & 0.946 {\scriptsize$\pm$.009} & 82.9\% \\
        & 0.15        & 15 & 0.988 {\scriptsize$\pm$.004} & 0.982 {\scriptsize$\pm$.005} & 49.4\% & 0.932 {\scriptsize$\pm$.010} & 58.7\% \\
        \midrule
        \textbf{$n_{daily} = 2500$}
        & 0.00 (Null) &  3 & 0.052 {\scriptsize$\pm$.009} & 0.020 {\scriptsize$\pm$.005} & 24.5\% & 0.947 {\scriptsize$\pm$.009} & --- \\
        & 0.05        &  3 & 0.368 {\scriptsize$\pm$.019} & 0.238 {\scriptsize$\pm$.017} & 10.0\% & 0.950 {\scriptsize$\pm$.009} & 63.6\% \\
        & 0.10 (MDE)  &  3 & 0.904 {\scriptsize$\pm$.012} & 0.830 {\scriptsize$\pm$.015} & 17.6\% & 0.988 {\scriptsize$\pm$.004} & 33.3\% \\
        & 0.15        &  3 & 0.997 {\scriptsize$\pm$.002} & 0.993 {\scriptsize$\pm$.003} & 34.5\% & 0.918 {\scriptsize$\pm$.011} & $-624\%$\,\footnotemark \\
        \bottomrule
    \end{tabular}
\end{table}
\footnotetext{Raw median bias is $0.0019$ and MUE median bias is $0.0135$ in this cell. The percentage-reduction metric is computed against a near-zero raw bias, mechanically inflating the magnitude. The substantive observation is that the MUE adjusts in the wrong direction, the same overcorrection regime documented in Section~\ref{sec:bias}.}

At the MDE, SPRT-z trails FHT power by 3.1 and 1.6 points at the finer cadences and 7.4 points at the coarsest. This shortfall occurs when $N_{FHT}$ spans only three enrolment days; the comparison does not isolate its cause. Sample reduction remains positive in every cell (10.0--54.3\%) and decreases as monitoring becomes coarser. Coverage is within Monte Carlo error of 95\% in six cells, including every null; four overcover, reaching 98.8\% at $(2500,0.10)$, and two $\delta=0.15$ cells undercover, reaching 91.8\% ($\pm1.1$). MUE bias reduction is 59--98\% in adequately powered fine-cadence cells, falls to 64\% and 33\% at the coarsest cadence, and reverses at $(2500,0.15)$.

The Brownian MUE and intervals can perform poorly when a test stops within a few looks, even though their simulation checks boundaries on a discrete grid. We therefore recommend against relying on either when the anticipated number of looks is small. In these experiments, this corresponds to roughly $N_{FHT}/n_{daily} \le 15$ with plausible effects at or above 1.5 times the MDE. Inversion that more closely reproduces the empirical accrual, variance estimates and prequential targets remains future work. A fixed-horizon interval at the stopped sample size is not a substitute, because sequential stopping invalidates its nominal coverage.

\bibliographystyle{plainnat}
\bibliography{references}

\end{document}